\documentclass[oneside]{elsart}
\usepackage{graphicx}
\usepackage{epstopdf}
\usepackage[figuresright]{rotating}
\usepackage{amsmath}
\usepackage{amssymb}
 
\def\intablecenterline#1{\hfill#1\hfill} 
\def\und{} 
\def\ETTunderline#1{\ifmmode\let\und=\underline\else 
\let\und=\underbar\fi\und{#1}}

\def\mathline{\ifmmode \vert\else$\vert$\fi}

\def\ETTstretch#1{\message{[[Sorry, you can't use Stretch with ETT]]}} 
 
\def\ETTexpand#1{\message{[[Sorry, you can't use Expand with ETT]]}} 
 
\def\xbackslash{\ifmmode \backslash\else$\backslash$\fi}

\def\quoteswitch{} 
\def\leftquotemarks{\ifmmode{}{}''\else``%
\let\quoteswitch=\rightquotemarks\fi} 
\def\rightquotemarks{\ifmmode{}{}''\else"\let\quoteswitch=\leftquotemarks\fi} 
\let\quoteswitch=\leftquotemarks 
\def\p$&-\${\overline{p}} 
\def\ETTstack#1#2{\ifmmode\ifx\int#1\displaystyle#1\limits#2\else 
\ifx\xint#1\displaystyle\int\limits#2\else 
\ifx\sum#1\displaystyle#1\limits#2\else%
\ifx\xsum#1\displaystyle\sum\limits#2\else 
\mathop{#1}\limits#2\fi\fi\fi\fi\else#1\mathop{#2}\limits\fi} 
 
\def\ETThalign#1{{\let\centerline=\intablecenterline 
\ifmmode\vcenter{\everycr={\noalign{\vskip1.5pt}}%
\halign{$\strut##$&$\strut##$&$\strut##$&$\strut##$&$\strut##$&%
$\strut##$&$\strut##$&$\strut##$&$\strut##$&$\strut##$&$\strut##$&%
$\strut##$&$\strut##$&$\strut##$&$\strut##$&$\strut##$&$\strut##$&%
$\strut##$&$\strut##$&$\strut##$\cr#1\crcr}\everycr={}}\else%
\hbox{$\vcenter{\everycr={\noalign{\vskip1.5pt}} 
\halign{\strut##&\strut##&\strut##&\strut##&%
\strut##&\strut##&\strut##&\strut##&\strut##&\strut##&\strut##&%
\strut##&\strut##&\strut##&\strut##&\strut##&\strut##&\strut##&%
\strut##&\strut##\cr#1\crcr}\everycr={}}$}\fi}} 
 
\gdef\ETTcolumn#1{\ifmmode%
\hbox{$\vcenter{\everycr{\noalign{\vskip-2pt}}%
\let\centerline=\intablecenterline 
\halign{\strut\hfill$##$\hfill\cr 
#1\crcr}\vskip-2pt}$}\else 
\hbox{$\vcenter{\offinterlineskip%
\let\centerline=\intablecenterline 
\halign{\strut\hfill##\hfill\cr 
#1\crcr}}$}\fi} 
 
\def\ETTtable#1{\ifmmode\vcenter{\everycr={\noalign{\vskip1.5pt}}%
\let\centerline=\intablecenterline%
\halign{\hfil$\vcenter{\hbox{$\strut##$}}$\hfil\ &\ %
\hfil$\vcenter{\hbox{$\strut##$}}$\hfil\ &\ %
\hfil$\vcenter{\hbox{$\strut##$}}$\hfil\ &\ %
\hfil$\vcenter{\hbox{$\strut##$}}$\hfil\ &\ %
\hfil$\vcenter{\hbox{$\strut##$}}$\hfil\ &\ %
\hfil$\vcenter{\hbox{$\strut##$}}$\hfil\ &\ %
\hfil$\vcenter{\hbox{$\strut##$}}$\hfil\ &\ %
\hfil$\vcenter{\hbox{$\strut##$}}$\hfil\ &\ %
\hfil$\vcenter{\hbox{$\strut##$}}$\hfil\ \cr 
#1\crcr}\everycr={}}\vrule depth3pt width0pt\relax\else 
\hbox{$\vcenter{\everycr={\noalign{\vskip1.5pt}}%
\let\centerline=\intablecenterline%
\halign{\hfil$\vcenter{\hbox{\strut##}}$\hfil\ &\ %
\hfil$\vcenter{\hbox{\strut##}}$\hfil\ &\ %
\hfil$\vcenter{\hbox{\strut##}}$\hfil\ &\ %
\hfil$\vcenter{\hbox{\strut##}}$\hfil\ &\ %
\hfil$\vcenter{\hbox{\strut##}}$\hfil\ &\ %
\hfil$\vcenter{\hbox{\strut##}}$\hfil\ &\ %
\hfil$\vcenter{\hbox{\strut##}}$\hfil\ &\ %
\hfil$\vcenter{\hbox{\strut##}}$\hfil\ &\ %
\hfil$\vcenter{\hbox{\strut##}}$\hfil\ \cr 
#1\crcr}\everycr={}}$}\vrule depth3pt width0pt\relax\fi} 
 
\def\ETTast{{\raise.25ex\hbox{$\ast$}}} 
 
\newif\ifinmath 
\newbox\superbox 
\newbox\superboxtwo 
 
\def\changetomath{$} 
 
\def\ETTsuperimpose#1#2{\ifmmode\let\check=\changetomath%
\else\let\check=\relax\fi%
\setbox\superbox=\hbox{\check#1\check}%
\setbox\superboxtwo=\hbox{\check#2\check}%
\ifdim\wd\superbox>\wd\superboxtwo%
\copy\superbox\hskip-\wd\superbox\hbox to 
\wd\superbox{\hfill\check#2\check\hfill}%
\else%
\copy\superboxtwo\hskip-\wd\superboxtwo\hbox to 
\wd\superboxtwo{\hfill\check#1\check\hfill}\fi} 
 
\def\hb{{\hskip3pt}} 
\let~=\hb

\newcount\moveup \newcount\moveover 
 
\def\lookagain{\futurelet\next\parser} 
 
\def\parser#1{\def\endb{}\ifx\next /\let\go=\relax \else\ifx\next r 
\global\advance\moveover by22\else 
\ifx\next l\global\advance\moveover by-22 
\else \ifx\next u\global\advance\moveup by22 
\else \ifx\next d\global\advance\moveup by-22\fi\fi\fi\fi 
\let\go=\lookagain\fi\go} 
 
\def\ETTadjust#1#2{\moveover=0 \moveup=0\setbox0=\hbox{\lookagain#1/}%
\divide\moveover by10 \divide\moveup by10\setbox1=\hbox{#2}%
\vtop to0pt{\hbox to\wd1{\hskip\moveover pt\raise\moveup pt\hbox{#2}\hss}%
\vss}} 
 
\newcount\upcounter 
\newcount\overcounter 
 
\newif\ifaz \newif\ifrz \newif\iflz \newif\ifdz \newif\ifuz 
 
\def\nonglobalreset{\azfalse\lzfalse\uzfalse%
\dzfalse\rzfalse}%
\long\gdef\ETTbox #1 #2{\def\parse{#1}%
\nonglobalreset\expandafter\looker\parse{}{}{}\boxer{\vbox{%
\let\centerline=\intablecenterline\halign{%
\strut##\cr#2\crcr}}}} 
 
\long\def\boxer#1{\ifaz\rztrue\lztrue\uztrue\dztrue\fi%
\hbox{$\,\,\vcenter{\ifuz\hrule\fi\hbox{\iflz\vrule\fi%
\hskip2pt\vbox{\vskip2pt\vbox{#1}\vskip2pt}%
\hskip2pt\ifrz\vrule\fi}\ifdz\hrule\fi}\,\,$}} 
 
\def\looker #1#2#3#4{%
\ifx#1A\aztrue\else\ifx#2a\aztrue\else
\ifx#1R\rztrue\else 
\ifx#2R\rztrue\else 
\ifx#3R\rztrue\else 
\ifx#4R\rztrue\fi\fi\fi\fi\fi 
\ifx#1L\lztrue\else 
\ifx#2L\lztrue\else 
\ifx#3L\lztrue\else 
\ifx#4L\lztrue\fi\fi\fi\fi 
\ifx#1U\uztrue\else 
\ifx#2U\uztrue\else 
\ifx#3U\uztrue\else 
\ifx#4U\uztrue\fi\fi\fi\fi 
\ifx#1D\dztrue\else 
\ifx#2D\dztrue\else 
\ifx#3D\dztrue\else 
\ifx#4D\dztrue\fi\fi\fi\fi\fi} 
 
\def\tie{\ifmmode\sim\else 
\hbox{\vtop to0pt{\hbox{\lower5pt\hbox{\~{}}}\vss}}\fi} 
\def\<{\ifmmode <\else $<$\fi} 
\def\>{\ifmmode >\else $>$\fi} 
 
\def\caret{\ifmmode{\vtop to0pt{\hbox{\lower7pt\hbox{$\hat{\vphantom{.}}$}} 
\vss}}\else \^{}\fi} 
 
\def\lcurlybracket{\ifmmode\{\else $\{$\fi} 
\def\rcurlybracket{\ifmmode\}\else $\}$\fi} 
 
\newdimen\tempdimen 
 
\def\smdarkP{\ \hbox{\global\setbox0=\hbox{\smallbackp}\vrule height\ht0 
depth\dp0 width0pt\vrule\hskip.6pt\vrule}\tempdimen=\ht0%
\advance\tempdimen by-.4pt%
\hskip-1pt\raise1.4pt\hbox{$\scriptstyle\bullet$}%
\hskip-2pt\llap{\raise\tempdimen\hbox to3pt{\hrulefill}}\ } 
 
\def\xxdarkP{\ \hbox{\global\setbox0=\hbox{\P}\vrule height\ht0 
depth\dp0 width0pt\vrule\hskip.6pt\vrule}\tempdimen=\ht0%
\advance\tempdimen by-.4pt\hskip-1pt\raise3pt\hbox{$\displaystyle\bullet$}%
\hskip-2pt\llap{\raise\tempdimen\hbox to4pt{\hrulefill}}\ }

\def\smallbackp{{\smallsymbol\char'173}}

\def\return{\hbox{\ \unskip{$\bf\leftarrow$\hskip-.5pt\raise2.4pt%
\hbox{\vrule height 2.5pt}}}}

\def\xxTM{\raise1ex\hbox{\amrseven TM}} 
 
\def\TMfive{\raise1ex\hbox{\amrfive TM}}

\def\xxeqcirc{\buildrel \lower1.5pt\hbox{$\scriptstyle\circ$}\over =} 
\def\smeqcirc{\buildrel \lower1.5pt\hbox{$\scriptscriptstyle\circ$}\over{\scriptstyle =}}

\def\xxeqdkcirc{\buildrel \lower1.5pt\hbox{$\scriptscriptstyle\bullet$}\over =} 
\def\smeqdkcirc{\buildrel \lower1.5pt\hbox{$\scriptscriptstyle\bullet$}\over{\scriptstyle =}}

\def\xxtridots{\ \unskip\raise4pt\hbox 
to1em{\hfil.\hfil}\llap{\hbox to1em{\hfil.\ .\hfil}}} 
 
\def\smtridots{\ \unskip\raise3pt\hbox 
to1em{\hfil.\hfil}\llap{\hbox to1em{\hfil.$\,$.\hfil}}}

\def\sqr#1#2{{\vcenter{\hrule height.#2pt 
\hbox{\vrule width.#2pt height#1pt\kern#1pt 
\vrule width.#2pt} 
\hrule height.2pt}}}

\def\xint{{\mathchoice{\int}{\displaystyle\int}{\int}{\int}}} 
\def\xsum{{\mathchoice{\sum}{\displaystyle\sum}{\sum}{\sum}}} 
 
\def\GermanS{\ifmmode\hbox{\ss}\else\ss\fi} 
\def\primeaccent{\ifmmode\hbox{\rm\char19}\else{\rm\char19}\fi} 
\def\underaccent{\ifmmode\hbox{\rm\char24}\else{\rm\char24}\fi} 
\def\EnglishPound{\ifmmode\hbox{\it\$}\,\else{\it\$}\fi} 
 
\newcommand{\be}{\begin{equation}}
\newcommand{\ee}{\end{equation}}
\newcommand{\bea}{\begin{eqnarray}}
\newcommand{\eea}{\end{eqnarray}}
\newcommand{\beas}{\begin{eqnarray*}}
\newcommand{\eeas}{\end{eqnarray*}}
\newcommand{\bi}{\begin{itemize}}
\newcommand{\ei}{\end{itemize}}
\newcommand{\bn}{\begin{enumerate}}
\newcommand{\en}{\end{enumerate}}

\def\lambar{{ \lambda \mkern-10mu\raise.5ex\hbox{--} }}

\def\thus{{ .. \mkern-7.5mu\raise.9ex\hbox{.} }\  }

\def\ba2#1#2{${\overline{#1}}^{#2}$}
\def\anti#1#2{\vbox{\ialign{##\crcr
     \hrulefill$\smash{\phantom{\scriptstyle#2}}$\crcr
     \noalign{\kern-1pt\nointerlineskip\vskip 0.25ex}
     $\hfil{#1}^{#2}\hfil$\crcr}}}
\def\anth#1#2{\vbox{\ialign{##\crcr
    \hrulefill$\smash{\phantom{\scriptstyle#2}}$\crcr
    \noalign{\kern-0.5pt\nointerlineskip\vskip 0.25ex}
    $\hfil{#1}^{#2}\hfil$\crcr}}}
\def\rless{{ r \mkern-.5mu\raise-.2ex\hbox{\tiny{<}} }}
\def\rmore{{ r \mkern-.5mu\raise-.2ex\hbox{\tiny{>}} }}

\newcommand{\Ket}{{\rm\bf\goodfontB Ket}}

\newcommand{\NOT}{{\rm\bf\goodfontB NOT }}

\newcommand{\QDENS}{{\rm\bf\goodfontB QDENSITY\,}}
\newcommand{\QCMPI}{{\rm\bf\goodfontB QCMPI\,}}

\def\goodfontB{\usefont{T1}{phv}{n}{n}\fontsize{9pt}{9pt}\selectfont}

\begin{document}
\begin{frontmatter}
\title{QCWAVE - A MATHEMATICA  QUANTUM COMPUTER SIMULATION UPDATE}

\author[pit]{Frank Tabakin}
\and
\author[spain]{Bruno Juli\'a-D\'{\i}az}

\address[pit]{Department of Physics and Astronomy \\
University of Pittsburgh, Pittsburgh,  PA, 15260}

\address[spain]{Departament de Estructura i Constituents de la Materia\\
Universitat de Barcelona, 08028 Barcelona (Spain)
}

\begin{abstract}
This Mathematica 7.0/8.0 package  upgrades and extends 
the quantum computer simulation code called QDENSITY.
Use of the density matrix was emphasized in QDENSITY, 
although that code was also applicable to a quantum state 
description.  In the present version, the quantum state 
version is stressed and made amenable to future extensions 
to parallel computer simulations. The add-on QCWAVE  extends 
QDENSITY in several ways. The first way is to describe 
the action of one, two and three- qubit quantum gates 
as a set of small ($2  \times 2, 4\times 4$ or $8\times 8$) 
matrices acting on the  $2^{n_q}$ amplitudes for a system of 
$n_q$ qubits.  This procedure was described in our parallel 
computer simulation QCMPI and is reviewed here. The advantage 
is that smaller storage demands are made, without loss of speed, 
and that the procedure can take advantage of message passing 
interface (MPI) techniques,  which will hopefully be generally 
available in future Mathematica versions.

Another extension of QDENSITY provided here is a multiverse 
approach, as described in our QCMPI paper. This multiverse 
approach involves using the present slave-master parallel 
processing capabilities of Mathematica 7.0/8.0 to simulate 
errors and error correction. The basic idea is that parallel 
versions of QCWAVE run simultaneously with random errors 
introduced on some of the processors, with an ensemble average 
used to represent the real world situation.  Within this approach, 
error correction steps can be simulated and their efficacy 
tested. This capability allows one to examine the detrimental 
effects of errors and the benefits of error correction on 
particular quantum algorithms.

Other upgrades provided in this version includes circuit-diagram 
drawing commands, better Dirac form and amplitude display features. 
These are included in the add-ons {\bf QCWave.m} and {\bf Circuits.m}, and 
are illustrated in tutorial notebooks.

In separate notebooks,  QCWAVE is applied to sample algorithms 
in which the parallel multiverse setup is illustrated and error 
correction is simulated. These extensions and upgrades will hopefully 
help in both instruction and in application to QC dynamics and 
error correction studies.

\end{abstract}
\end{frontmatter}

\newpage
\noindent{\bf Program Summary}

{\it Title of program:} QCWAVE. \
{\it Catalogue identifier:}\\
{\it Program summary URL:} http://cpc.cs.qub.ac.uk/summaries\\
{\it Program available from:} CPC Program Library, Queen's University of Belfast, N. Ireland. \\
{\it Operating systems:} Any operating system that supports Mathematica;
tested under Microsoft Windows XP, Macintosh OSX, and Linux FC4.\\
{\it Programming language used:} Mathematica 7.0. \\
{\it Number of bytes in distributed program, including test code and 
documentation: xx}\\
{\it Distribution format:} tar.gz\\
{\it Nature of Problem:} Simulation of quantum circuits, 
quantum algorithms, noise and quantum error correction.\\
{\it Method of Solution:} A Mathematica package containing commands to create 
and analyze quantum circuits is upgraded and extended, with emphasis on state 
amplitudes. Several Mathematica notebooks containing relevant examples are 
explained in detail. The parallel computing feature of Mathematica is used to 
develop a multiverse approach for including noise and forming suitable
ensemble averaged density matrix evolution. Error correction is simulated.\\

\newpage

\section{INTRODUCTION}

In this paper, QDENSITY~\cite{QDENSITY} (a Mathematica~\cite{Mathematica} 
package that  provides a flexible simulation of a quantum computer) is
extended and upgraded by an add-on called QCWAVE~\footnote{Other  
 authors  have also  developed  Mathematica/Maxima QDENSITY  based 
 quantum computing simulations~\cite{mexico,qinf}. Hopefully, 
they will incorporate the ideas we provide herein in their future efforts.}. 
The earlier flexibility in QDENSITY is enhanced by adopting a simple state 
vector approach to initializations, operators, gates, and measurements. 
Although the present version stresses a state vector approach the 
density matrix can always be constructed and examined. Indeed, a 
parallel universe (or multiverse) approach is also included, using the 
present Mathematica 7.0/8.0 slave-master feature. This multiverse approach, 
which was published~\cite{QCMPI} in our QCMPI paper\!~\footnote{\QCMPI is 
a quantum computer (QC) simulation package written in Fortran 90 with 
parallel processing capabilities.}, allows separate dynamical evolutions 
on several processors with some evolutions subject to random errors. Then 
an ensemble average is performed over the various processors to produce 
a density matrix that describes a QC system with realistic errors. Error 
correction methods can also be invoked on the set of processors to test 
the efficacy of such methods on selected QC algorithms.

In section~\ref{sec2}, we introduce qubit state vectors and associated
amplitudes for one, two and multi-qubit states. In section~\ref{sec3}, 
a method for handling one, two and three- qubit operators acting on 
state vectors with commands from QCWAVE are presented.  

In section~\ref{sec4}, illustrations of how to apply gates to states are 
shown. In section~\ref{sec5}, the multiverse approach is described and 
the parallel method for introduction of errors and error correction are 
given. The ensemble averaged density matrix is then constructed. 

Additional upgrades, such as Dirac and amplitude displays and circuit 
drawing are presented in section~\ref{sec6}. Suggested applications are 
presented in the conclusion section~\ref{sec7}.

\section{MULTI-QUBIT STATES}
\label{sec2}

\subsection{One-qubit states}
\label{sec2a}

The basic idea of a quantum state, its representation in Hilbert space and 
the concepts of quantum computing have been discussed in many 
texts~\cite{DIRAC,MESSIAH,Nielsen}. A brief review was given in our earlier 
papers in this series~\cite{QDENSITY,QCMPI}.  Here we proceed directly 
from one, two  and multi-qubit states and their amplitudes to how 
various operators alter those amplitudes.

To start, recall that when one focuses on just two states of a quantum 
system, such as the spin part of a spin-1/2 particle, the two states 
are represented as either $\mid 0 \rangle$  or $\mid 1\rangle.$
A one qubit state is a superposition of the two states associated with 
the above $0$ and $1$ bits: 
\be 
\mid \Psi_1\rangle =C_0  \mid 0\rangle+C_ 1  \mid 1\rangle, 
\ee 
where $C_0 \equiv \langle0\mid \Psi_1\rangle$ and 
$C_1 \equiv \langle1\mid \Psi_1\rangle$ are complex
probability amplitudes for finding the qubit in the state $\mid0\rangle$ 
or  $\mid1\rangle ,$ respectively. The normalization of the state
$\langle\Psi_1\mid\Psi_1\rangle =1$, yields $\mid C_0 \mid^2 + \mid C_1 \mid^2=1$.
Note that the spatial aspects of the wave function are being suppressed; 
which corresponds to the particle being in a fixed location. 
The kets $\mid 0\rangle$ and $\mid 1\rangle$ can be represented as
$  \mid 0\rangle \rightarrow
   \left(
\begin{smallmatrix}
1 \\
 0
\end{smallmatrix}
\right)  $ 
and
$ \mid 1\rangle \rightarrow
   \left(
\begin{smallmatrix}
0 \\
 1
\end{smallmatrix}
\right).  $
Hence a $2 \times 1$ matrix representation of this one-qubit state is:
$
\mid \Psi_1 \rangle \rightarrow  \left(
\begin{smallmatrix}
 C_0 \\
 C_1
\end{smallmatrix}
\right)    \, .   
$ 

An essential point is that a quantum mechanical (QM) system can exist 
in a superposition of these two bits; hence, the state is called a
quantum-bit or ``qubit''. Although our discussion uses the notation
of a system with spin 1/2, it should be noted that the same discussion
applies to any two distinct quantum states that can be associated 
with $\mid 0\rangle $ and $ \mid 1\rangle$.

\subsection{Two-qubit states}
\label{sec2b}

The single qubit case can now be generalized to multiple qubits. Consider 
the product space of two qubits both in the ``up" $\mid 0\rangle $ state 
and denote that product state as $ \mid 0\ 0\rangle= \mid 0\rangle \mid
0\rangle, $ which clearly generalizes to
\be
\mid q_1\ q_2\rangle= \mid q_1\rangle \mid q_2\rangle,
\ee
where $q_1,q_2$ in general take on the values $0$ and $1$. This product 
is called a tensor product and is symbolized as
\be
\mid q_1\ q_2\rangle= 
\mid q_1\rangle \otimes \mid q_2\rangle. 
\ee   

In \QDENS, the kets $\mid0\rangle,\mid1\rangle$ are invoked by the commands
$\Ket[0]$ and $\Ket[1],$ and the product state by for example 
$\mid 00\rangle= \Ket[0] \otimes  \Ket[0] .$

The kets $ \mid 0 0\rangle , \mid 01\rangle,  \mid 10\rangle,  \&  \mid
11\rangle $ can be represented as $4 \times 1$  matrices
\be
 \mid 00\rangle \rightarrow
   \left(
\begin{array}{l}
1 \\
 0 \\
 0 \\
 0
\end{array}
\right) ; 
 \mid 01\rangle \rightarrow
   \left(
\begin{array}{l}
0 \\
 1 \\
 0 \\
 0
\end{array}
\right) ;
 \mid10\rangle \rightarrow
   \left(
\begin{array}{l}
0 \\
 0 \\
 1 \\
 0
\end{array}
\right)  ;
 \mid11\rangle \rightarrow
   \left(
\begin{array}{l}
0 \\
 0 \\
 0 \\
 1
\end{array}
\right).  
\ee
Hence, a $4 \times 1$ matrix representation of the two-qubit 
state
\be 
\mid \Psi_2\rangle =C_0  \mid 00\rangle+C_ 1  \mid 0 1\rangle +C_ 2  \mid 1
0\rangle +C_ 3  \mid 1 1\rangle, 
\ee 
is:
\be
\mid \Psi_2 \rangle \rightarrow  \left(
\begin{array}{l}
 C_0 \\
  C_1 \\
   C_2 \\
 C_3
\end{array}
\right)    \, .   
\label{amp2} 
\ee 
Again  
$C_0\equiv \langle 00\mid \Psi_2\rangle, 
C_1\equiv \langle 01\mid \Psi_2\rangle, 
C_2\equiv \langle 10\mid \Psi_2\rangle,$ and
$C_3\equiv \langle 11\mid \Psi_2\rangle,$
 are complex probability amplitudes for finding the two-qubit 
system in the states  $\mid q_1\ q_2\rangle.$  The normalization of the 
state $\langle\Psi_2\mid\Psi_2\rangle =1$, yields 
\be 
\mid C_0 \mid^2 + \mid C_1 \mid^2+ \mid C_2 \mid^2+ \mid C_3 \mid^2=1. 
\ee
Note that we label the amplitudes using the decimal equivalent of 
the bit product $q_1\ q_2,$ so that for example a binary label on 
the amplitude $ C_{ 1 0} $ is equivalent to the decimal label $C_2 .$

\subsection{Multi-qubit states}
\label{sec2c}

For $n_q$ qubits the computational basis of states generalizes to:
 \be
\mid n \rangle_{n_q} \equiv \mid q_1\rangle  \cdots  \mid q_{n_q}\rangle\equiv
\mid q_1\ q_2\ \cdots\ q_{n_q}\rangle \equiv \mid {\bf Q} \rangle .
\ee
We use the convention that the most significant qubit is labeled as $q_1$ 
and the least significant qubit by $q_{n_q}.$ Note we use $q_{i}$ to indicate 
the quantum number of the $i$th qubit. The values assumed by any qubit is 
limited to either $q_i = 0$ or $1.$  The state label ${\bf Q}$  denotes the 
qubit array ${ \bf{Q}} = \left(  q_1, q_2, \cdots,   q_{n_q} \right) ,$
which is a binary number label for the state with equivalent decimal label $n.$ 
This decimal multi-qubit state label is related to the equivalent binary label by 
\be
 n  \equiv  q_1 \cdot  2^{n_q -1}+ q_{2} \cdot   2^{n_q -2} 
+ \cdots+  q_{n_q} \cdot   2^{0} = \sum_{i=1}^{n_q} \, q_i  \cdot   2^{n_q -i} \, .
\label{staten}
\ee  
Note that the $i$th qubit contributes a value of $q_i \cdot   2^{n_q -i}$ to 
the decimal number $n.$  Later we will consider ``partner states'' 
($ \mid { n_0}   \rangle,\   \mid { n_1} \rangle $) associated with a given 
${ n}, $ where a particular qubit  $i_s$  has a value of $q_{i_s} = 0,$
\be
{ n_0} =  { n} - q_{i_s}   \cdot   2^{n_q -i_s},
\label{pair0}
\ee 
or a value  of
$q_{i_s} = 1,$
\be
{ n_1} =  { n} -  (q_{i_s}-1)   \cdot   2^{n_q -i_s}.
\label{pair1}
\ee  
These partner states are involved in the action of a single operator acting 
on qubit  $i_s ,$ as described in the next section.

A general state with $n_q$ qubits can be expanded in terms of the above 
computational basis states as follows
\be
\mid \Psi\rangle_{n_q} = \sum_{ \bf Q} C_{ \bf Q} \mid { \bf Q} \rangle 
\equiv   \sum_{ n=0}^{2^{n_q}-1 }  C_{ n} \,  \mid  n \rangle\,,
\ee  
where the sum over ${ \bf Q}$ is really a product of $n_q$ summations 
of the form $\sum_{q_i=0,1}.$ The above Hilbert space expression maps over 
to an array, or column vector, of length $2^{n_q}$
\bea
  \qquad  \mid \Psi\rangle_{n_q}&\equiv& \left(
\begin{array}{l}
 C_0 \\
 C_1 \\
\ \, \vdots \\
\ \, \vdots \\
  C_{2^{n_q} -1}
\label{ampn}\end{array}
\right)   
\qquad
{\rm or\ with\ binary\ labels}
 \longrightarrow  
\left(
\begin{array}{lc}
 C_{0 \cdots 00} \\
 C_{0 \cdots 01} \\
\ \, \vdots \\
\ \, \vdots \\
 C_{1 \cdots 11}
\end{array}\right) \, .
 \eea    
The expansion coefficients $C_n$ (or $C_{{\bf Q}}$)  are complex numbers 
with the physical meaning that $C_n=\langle n \mid \Psi\rangle_{n_q}$ is 
the probability amplitude for finding the system in the computational 
basis state $\mid n\rangle, $ which corresponds to having the qubits 
pointing in the directions specified by the binary array ${\bf Q}.$
Switching between decimal $n$ and equivalent binary ${\bf Q}$ labels is
accomplished by the Mathematica command {  \bf IntegerDigits}.      

In general, the complex amplitudes $C_n$ vary with time and are changed 
by the action of operators or gates, as outlined next.

\section{MULTI-QUBIT OPERATORS}
\label{sec3}

Operators that act in the multi-qubit space described above can be 
generated from a set of separate Pauli operators~\footnote{The Pauli 
operators act in the qubit Hilbert space, and have the matrix representation:
 $\sigma_x = \left(
\begin{smallmatrix}
0& 1\\
1& 0
\end{smallmatrix}\right) \  ; \ 
\sigma_y = \left(
\begin{smallmatrix}
0& -I\\
I& 0
\end{smallmatrix}\right) \  ; \ 
\sigma_z = \left(
\begin{smallmatrix}
1& 0\\
0& -1
\end{smallmatrix}\right)$.   
Here $I \equiv \sqrt{ -1}\ .$
}, 
acting in each qubit space. These separate Pauli operators refer to distinct 
quantum systems and hence they commute. Note, Pauli operators acting on 
the same qubit do not commute; indeed, they have the property 
$\sigma_i  \sigma_j  -  \sigma_j  \sigma_i =2
\  i\   \epsilon_{ijk}\  \sigma_k.$ 
The Pauli operator  $\sigma_0$ is just the unit $2\times 2$ matrix. 
We denote a Pauli operator acting on qubit $i_s$ as 
$\sigma_k^{ (i_s)},$ where $k=(x,y,z)=(1,2,3)$ is the component 
of the Pauli operator. For example, the tensor product of two qubit 
operators  has the following structure
\bea  
\langle  a_1 \mid \sigma_i \mid b_1\rangle  \langle  a_2 \mid \sigma_j \mid b_2\rangle&=&
 \langle  a_1 a_2  \mid \sigma^{(1)}_i \sigma^{(2)}_j \mid b_1 b_2\rangle \nonumber \\
&=&
 \langle  a_1 a_2  \mid \sigma^{(1)}_i  \otimes  \sigma^{(2)}_j \mid b_1 b_2\rangle\,,
\eea
which defines what we mean by the tensor product of two qubit operators 
$\sigma^{(1)}_i \otimes \sigma^{(2)}_j.$ The generalization to more qubits 
is immediate
\be
( \sigma^{(1)}_i  \otimes  \sigma^{(2)}_j )  \otimes( \sigma^{(3)}_k
\otimes  \sigma^{(4)}_l) \cdots \  .
\ee

\subsection{One-qubit operators}
\label{sec3a}

One-qubit operators change the amplitude coefficients of the quantum state.  
The \NOT and Hadamard ${\bf \cal{ H} }$  are examples of one-qubit operators 
of particular interest:
$
\NOT \equiv\sigma_x=\left(
\begin{smallmatrix}
0& 1\\
1& 0
\end{smallmatrix}\right) \, , 
 {\bf \cal{ H} }\equiv \frac{ \sigma_x + \sigma_z}{\sqrt{2}}
= \frac{1}{ \sqrt{2}}  \left(
\begin{smallmatrix}
1& 1\\
1& -1
\end{smallmatrix}\right)   \, .
\label{NOTH}  $    
These have the following effect on the basis states 
$\NOT \mid 0\rangle  = \mid 1\rangle$, $\NOT \mid 1\rangle  = \mid 0\rangle,$ 
and  
${\bf \cal{ H} }  \mid 0\rangle = \frac{ \mid 0\rangle +  \mid
  1\rangle}{\sqrt{2}}$, and 
${\bf \cal{ H} }  \mid 1\rangle = \frac{ \mid 0\rangle -  \mid 1\rangle}{\sqrt{2}}.$
 
General one-qubit operators can also be constructed from the Pauli 
operators;  we denote the general one-qubit operator acting on qubit $s$ 
as $ {\Omega_s }.$  Consider the action of such an operator on the multi-qubit 
state $\mid \Psi\rangle_{n_q} : $
 \bea
 {\Omega_s }\!\!  \mid \Psi\rangle_{n_q}&=& \sum_{{\bf Q} }    C_{{\bf Q} } \  \      {\Omega_s }\!
 \!  \mid  {{\bf Q}}  \rangle \\ \nonumber
 &=&
\sum_{q_1=0,1}   \cdots  \sum_{q_s=0,1}   \cdots   \sum_{q_{n_q}=0,1}\ C_{{\bf Q} } \  \      
\mid  q_1 \rangle  \cdots \  (  {\Omega_s }\!\! \mid  q_s  \rangle )\    \cdots  \mid q_{n_q} \rangle. \\
\label{op1}
\eea  
Here $ {\Omega_s }$ is assumed to act only on the qubit $i_s$ of value 
$q_s.$  The $(  {\Omega_s }\!\! \mid  q_s  \rangle )$ term can be expressed 
as
\be
    {\Omega_s }\!\! \mid  q_s  \rangle =  \sum_{q'_s=0,1}\!\!   
\mid q'_s \rangle \langle q'_s \mid\! {\Omega_s }\! \mid  q_s  \rangle,
 \label{closure1} 
\ee  
using the closure property of the one qubit states. Thus Eq.~(\ref{op1}) becomes
\bea  
 {\Omega_s} \!\!  \mid \Psi\rangle_{n_q}&=& \sum_{\bf  Q}   C_{\bf Q}\   {\Omega_s}  \!\! \mid  { \bf Q } \rangle
=  \\   \nonumber
\sum_{q_1=0,1}   \cdots  \sum_{q_s=0,1}   \cdots \!\!   \sum_{q_{n_q}=0,1}  \sum_{q'_s=0,1} \!\! C_{\bf Q} \!\!    
&&\langle q'_s \mid \!{\Omega_s }\! \mid  q_s  \rangle\  \mid  q_1 \rangle  \cdots \   \mid  q'_s  \rangle  \cdots  \mid q_{n_q} \rangle.
\eea  
Now we can interchange the labels $ q_s \leftrightarrow  q'_s ,$ and use 
the label $ {\bf Q} $ to obtain the algebraic result for the action of a 
one-qubit operator on a multi-qubit state
\be
{\Omega_s }  \mid \Psi\rangle_{n_q}= \sum_{\bf  Q}   {\tilde C}_{\bf Q}\   \mid  { \bf Q } \rangle =
 \sum_{n=0}^{ 2^{n_q}-1  }    {\tilde C}_{n}\   \mid n\rangle,
\ee    
where
\be
{\tilde C}_{\bf Q}=  {\tilde C}_{n} =    \sum_{q'_s=0,1}\!\! 
\langle q_s \mid \!{\Omega_s }\! \mid  q'_s  \rangle\ C_{\bf Q' ,} 
\label{oneopres}
\ee  
where  $ { \bf{Q}} = \left(  q_1, q_2, \cdots   q_{n_q} \right) ,$ and  
$ { \bf{Q'}} = \left(  q_1,\cdots q'_{s} \cdots  q_{n_q} \right) .$  
That is ${ \bf Q}$  and ${ \bf Q'}$ are equal except for the qubit 
acted upon by the one-body operator  ${\Omega_s}.$  
    
A better way to state the above result is to consider Eq.~(\ref{oneopres}) 
for the case that $n$ has $q_s=0$ and thus $n\rightarrow n_0$ and to write 
out the sum over $q'_s$ to get
\be
{\tilde C}_{n_0} = \langle 0 \mid \!{\Omega_s }\! \mid  0  \rangle C_{n_0}+\langle 0  \mid \!{\Omega_s }\! \mid  1  \rangle C_{n_1},
\ee  
where we introduced the partner to $n_0$ namely $n_1.$ For the case that 
$n$ has $q_s=1$ and thus $n\rightarrow n_1$ Eq.~(\ref{oneopres}), with 
expansion of the sum over $q'_s$ yields 
\be
      {\tilde C}_{n_1} = \langle 1 \mid \!{\Omega_s }\! \mid  0  \rangle C_{n_0}+\langle 1
      \mid \!{\Omega_s }\! \mid  1  \rangle C_{n_1}. 
\ee 
or written as a matrix equation we have for each $n_0, n_1$ partner pair
 \be
  \left(
\begin{array}{l}
  {\tilde C}_{n_0} \\
  {\tilde C}_{n_1}
\end{array}
\right)   
=  \left(
\begin{array}{lccr}
\langle 0 \mid {\Omega_s} \mid  0  \rangle&&& \langle 0 \mid{\Omega_s} \mid  1  \rangle\\
\langle 1 \mid {\Omega_s} \mid  0  \rangle&&&\langle 1 \mid {\Omega_s} \mid  1  \rangle
\end{array}  \right) 
 \left(
\begin{array}{l}
 C_{n_0} \\
 C_{n_1}
\end{array}
\right)
 \label{resmat1} 
\ee 
This is not an unexpected result.       
    
Equation~(\ref{resmat1}) above shows how a $2\times2$ one-qubit operator 
${\Omega_s}$ acting on qubit $i_s$ changes the state amplitude for each 
value of $n_0.$  Here, $n_0$ denotes a decimal number for a computational 
basis state with qubit $i_s$ having the $q_s$ value zero and $n_1$ denotes 
its partner decimal number for a computational basis state with qubit $i_s$ 
having the $q_s$ value one. They are related by
\be
n_1 = n_0 + 2^{n_q-i_s}.
\ee   
At times, we shall call $ 2^{n_q-i_s}$ the ``stride'' of the $i_s$ 
qubit;  it is the step in $n$ needed to get to a partner. There are 
$2^{n_q}/2$ values of $n_0$ and hence $2^{n_q}/2$ pairs $n_0,n_1.$ 
Equation ~(\ref{resmat1}) is applied to each of these pairs. 
In QCWAVE that process is included in the command 
{\bf Op1}~\footnote{{\bf Op1} yields result of a one-body operator 
${\Omega}$ acting on qubit $``is'"$ in state $\psi_0$; the result 
is the final state $\psi_f$. Called as: $\psi_f= {\bf Op1}[ \Omega, is ,\psi_0].$
}
  
Note that we have replaced the full $2^{n_q}\times2^{n_q}$ one qubit 
operator by a series of $2^{n_q}/2$ sparse  matrices. Thus we do not 
have to store the full  $2^{n_q}\times2^{n_q}$ but simply provide a 
$2\times2$ matrix for repeated use. Each application of the $2\times2$ 
matrix involves distinct amplitude partners and therefore the set of 
$2\times2$ operations can occur simultaneously and hence in parallel. 
That parallel advantage was employed in our QCMPI fortran version, using 
the MPI~\cite{MPI} protocol for inter-processor communication.  The 7.0 \& 8.0 
versions of Mathematica include only master-slave communication and 
therefore this advantage is not generally available. It is possible 
to use MPI with Mathematica~\cite{pooch},  but only at considerable 
cost.  Another promising idea is to use the ``CLOJURATICA''~\cite{CLOJURATICA} 
package, but that entails an additional language. So the full MPI advantage 
will have to wait until MPI becomes available hopefully on future 
Mathematica versions.
    
In the next section, this procedure is generalized to two- and three-qubit 
operators,  using the same concepts.
   
\subsection{Two-qubit operators}
\label{sec3b}
The case of a two-qubit operator is a generalization of the steps 
discussed for a one-qubit operator. Nevertheless, it is worthwhile 
to present those details, as a guide to those who plan to use and 
perhaps extend QCWAVE.

We now consider a general two-qubit operator that we assume acts 
on qubits $i_{s_1}$ and $i_{s_2},$  each of which ranges over the 
full ${1,  \cdots,  n_q}$ possible qubits. General two-qubit operators 
can be constructed from tensor products of two Pauli operators;  
we denote the general two-qubit operator as $ {\bf \cal{ V} }.$  
Consider the action of such an operator on the multi-qubit 
state $\mid \Psi \rangle_{n_q} : $
\bea
{\bf  \cal{ V} }\!\!  \mid \Psi\rangle_{n_q}&=& \sum_{{\bf Q} }    C_{{\bf Q} } \  \      {\bf \cal{ V} }\!
 \!  \mid  {{\bf Q}}  \rangle   \\ \nonumber
 &=&
\sum_{q_1=0}^{1}   \cdots  \sum_{q_{s1},q_{s2}=0}^{1}    \cdots  \sum_{q_{n_q}=0}^{1}\ C_{{\bf Q} } \  \        
\mid  q_1 \rangle  \cdots  (  {\bf {\cal V} }\!\! \mid  q_{s1}  q_{s2}   \rangle )\    \cdots  \mid q_{n_q} \rangle.
\label{op2}
\eea  
  
Here $ {\bf \cal{ V} }$ is assumed to act only on the two $q_{s1},q_{s2}$ 
qubits. The $(  {\bf {\cal V} }\!\! \mid  q_{s1}\    q_{s2}   \rangle  )$ term 
can be expressed as
\be
{\bf {\cal V} }\!\! \mid  q_{s1}\  q_{s2}   \rangle  
=  \sum_{q'_{s1},q'_{s2}=0}^{1}  \mid q'_{s1}\  q'_{s2} \rangle \langle
q'_{s1}\  q'_{s2}  \mid\! {\bf {\cal V} }\!\mid  q_{s1}\   q_{s2}   
\rangle 
\label{closure2} 
\ee  
using the closure property of the two-qubit product states. 
Thus Eq.~(\ref{op2}) becomes
\bea  
 {\bf { \cal V} } \!\!  \mid \Psi\rangle_{n_q}&=& \sum_{\bf  Q}   C_{\bf Q}\   {\bf {\cal V}}  \!\! \mid  { \bf Q } \rangle
= 
\sum_{q_1=0}^{1}   \cdots  \sum_{q_{s1}=0}^{1}   \cdots     \sum_{q_{s2}=0}^{1}   \cdots    \sum_{q_{n_q}
=0}^{1}  \sum_{q'_{s1},q'_{s2}=0}^{1}    \\   \nonumber C_{\bf Q}    
&&\langle q'_{s1} q'_{s2}  \mid \!{\bf {\cal V} }\! \mid  q_{s1} q_{s1}
\rangle\  \mid  q_1 \rangle  
\cdots \   \mid  q'_{s1}     q'_{s2}  \rangle \cdots \   \mid q_{n_q} \rangle.
\eea  
Now we can interchange the labels 
$ q_{s1} \leftrightarrow  q'_{s1} , q_{s2} \leftrightarrow  q'_{s2} $  
and use the label $ {\bf Q} $ to obtain the algebraic result for the 
action of a two-qubit operator on a multi-qubit state,
 
\be
{\bf  \cal{ V} }  \mid \Psi\rangle_{n_q}= \sum_{\bf  Q}  
 {\tilde C}_{\bf Q}\   \mid  { \bf Q } \rangle =
\sum_{n=0}^{ 2^{n_q}-1  }    {\tilde C}_{n}\   \mid n\rangle,
\ee    
where
\be
{\tilde C}_{\bf Q}=  {\tilde C}_{n} =    \sum_{q'_{s1},q'_{s2}=0}^{1}\!\!
\langle q_{s1}  q_{s2} \mid \!{\Omega_s }\! \mid  q'_{s1}  q'_{s2}  \rangle\ 
C_{\bf Q' ,}
\label{twoopA}
\ee  
where 
${ \bf{Q}} = \left(  q_1, q_2, \cdots   q_{n_q} \right) ,$ and  
$ { \bf{Q'}} = \left(  q_1,\cdots  q'_{s1} \cdots q'_{s2} \cdots q_{n_q}
\right) .$ That is ${ \bf Q}$  and ${ \bf Q'}$ are equal except for 
the qubits acted upon by the two-body operator ${\bf {\cal V}}.$  
    
A better way to state the above result is to consider 
Eq.~(\ref{twoopA}) for the following four choices 
\bea
 n_{ 00} & \rightarrow &  ( q_1 \cdots   q_{s1}=0  \cdots  q_{s2}=0 ,\cdots   q_{n_q} ) \nonumber \\
 n_{01}   &  \rightarrow  & ( q_1 \cdots   q_{s1}=0  \cdots  q_{s2}=1 ,\cdots   q_{n_q} ) \nonumber \\
 n_{10}   & \rightarrow &  ( q_1 \cdots   q_{s1}=1  \cdots  q_{s2}=0 ,\cdots   q_{n_q} ) \nonumber \\
 n_{11}   & \rightarrow &  ( q_1 \cdots   q_{s1}=1  \cdots  q_{s2}=1 ,\cdots   q_{n_q} ) , 
\label{twoopB}
\eea
where the computational basis state label $n_{q_{s1},q_{s2}}$ denotes 
the four decimal numbers  corresponding to ${\bf Q} = ( q_1, \cdots q_{s1} 
\cdots  q_{s2 }  \cdots  q_{n_q}).$  
     
Evaluating Eq.~(\ref{twoopA}) for the four choices Eq.~(\ref{twoopB}) 
and completing the sums over $q'_{s1}, q'_{s2} ,$  the effect of a general
two-qubit operator on a multi-qubit state amplitudes is given by a 
$4 \times 4$ matrix
    
\be
\left(
\begin{array}{l}
  {\tilde C}_{n_{00}} \\
  {\tilde C}_{n_{01}} \\
  {\tilde C}_{n_{10} }\\
  {\tilde C}_{n_{11}}
\end{array}
\right)   
=  \left(
\begin{array}{lccccr}
 {\cal V}_{00;00} &&  {\cal V}_{00;01}  
 \ \ \  {\cal V}_{00;10}      \ \ \       {\cal V}_{00;11}\\
{\cal V}_{01;00} &&  {\cal V}_{01;01}  
 \ \ \  {\cal V}_{01;10}      \ \ \       {\cal V}_{01;11}\\
 {\cal V}_{10;00} &&  {\cal V}_{10;01}  
 \ \ \  {\cal V}_{10;10}      \ \ \       {\cal V}_{10;11}\\
{\cal V}_{11;00} &&  {\cal V}_{11;01}  
 \ \ \  {\cal V}_{11;10}      \ \ \       {\cal V}_{11;11}
\end{array}  \right) 
 \left(
\begin{array}{l}
 C_{n_{00}} \\
  C_{n_{01}} \\
  C_{n_{10} }\\
 C_{n_{11}}
\end{array} 
\right) \ ,
\label{resmat2} 
\ee   
where $ {\cal V}_{ij;kl} \equiv  \langle i,j   \mid  {\cal V} \mid  k , l    \rangle.$
Equation~(\ref{resmat2}) shows how a $4\times4$ two-qubit operator ${\cal V}$ 
acting on qubits  $i_{s1},i_{s2} $ changes the state amplitude for each value 
of $n_{00}.$ Here, $n_{00}$ denotes a decimal number for a computational basis 
state with qubits $i_{s1},i_{s2} $ both having the values zero and its three 
partner decimal numbers for a computational basis state with qubits
$i_{s1},i_{s2} $ having the values $(0,1), (1,0)$ and $(1,1),$ respectively. 
The four partners $n_{00},n_{01},n_{10},n_{11},$ or ``amplitude quartet'',
coupled by the two-qubit operator are related by:
\be
n_{01}=n_{00} +  2^{n_q - i_{s2}} \qquad n_{10}=n_{00} +  2^{n_q - i_{s1}} \qquad n_{11}=n_{00} +  2^{n_q - i_{s1}}+2^{n_q - i_{s2}},
\ee  
where $ i_{s2}, i_{s2}$ label the quarks that are acted on by the 
two-qubit operator.   
  
There are $2^{n_q}/4$ values of $n_{00}$ and hence $2^{n_q}/4$ amplitude 
quartets $n_{00}, n_{01}, n_{10}, n_{11}.$ Equation~(\ref{resmat2}) is 
applied to each of these quartets for a given pair of struck qubits. 
In QCWAVE that process is included in the command 
{\bf Op2}~\footnote{{\bf Op2} yields result of a two-body operator 
${\Omega}$ acting on qubits $``is"$ and $``is2"$  in state $\psi_0$; 
the result is the final state $\psi_f$. Called as: 
$\psi_f=  {\bf Op2}[ \Omega, is1,is2 ,\psi_0].$}.
  
In this treatment, we are essentially replacing a large sparse 
matrix, by a set of  $2^{n_q}/4$ $4 \times 4 $ matrix actions, thereby 
saving the storage of that large matrix.

\subsection{Three-qubit operators}
\label{sec3c}
The above procedure can be extended to the case of three-qubit operators.  
Instead of pairs or quartets of states that are modified, we now have an 
octet of states modified by the three-qubit operator and  $2^{n_q}/8$ repeats 
to cover the full change induced in the amplitude coefficients. For brevity 
we omit the derivation. In QCWAVE that process has been implemented by the 
command {\bf Op3}~\footnote{ {\bf Op3} yields result of a three-body operator 
${\Omega}$ acting on qubits $``is1"$,$``is2"$ and $``is3"$  in state $\psi_0$; 
the result is the final state $\psi_f$. Called as: 
$\psi_f={\bf Op3}[ \Omega, is1,is2 ,is3,\psi_0].$}

\section{IMPLEMENTATION OF GATES ON STATES}
\label{sec4}

We now present some sample cases in which the above gates are applied to 
state vectors.

\subsection{One-qubit operators}
\label{sec4a}

\begin{figure}[t]
\begin{center}
\fbox{\parbox{.5\textwidth}{\includegraphics[width=.5\textwidth]{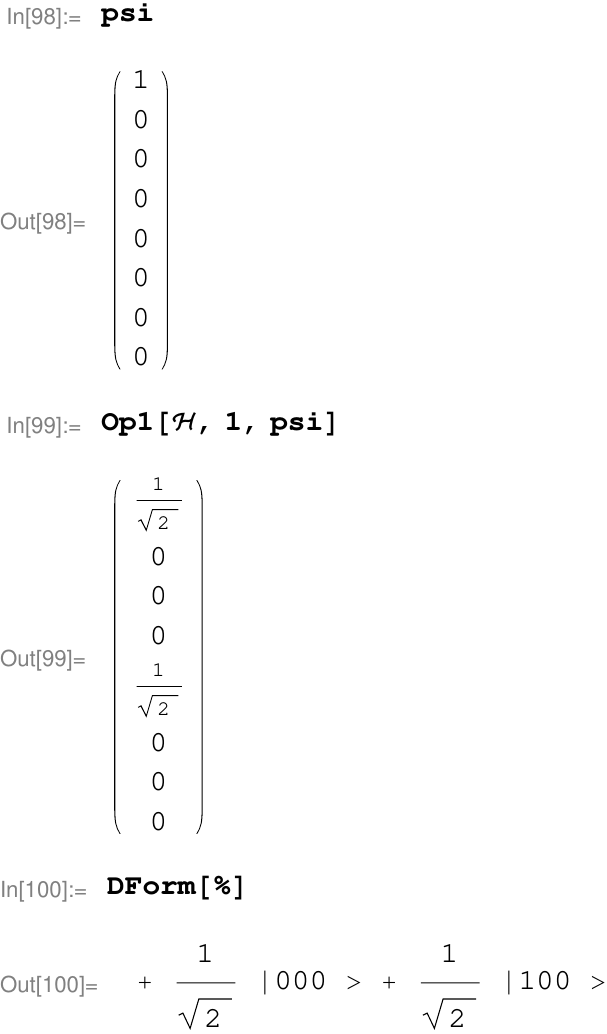}}}
 \caption{One Hadamard example. Here psi$= \mid 0 0 0 \rangle.$}
\protect\label{OneHad}
\end{center}
\end{figure}

Consider a state vector for $n_q=3$ qubits defined by $|\Psi\rangle_3= \mid  0 0 0\rangle,$
which in vector form is
\be
\mid \Psi_3 \rangle =  \left(
\begin{array}{l}
 C_0 \\
  C_1 \\
   C_2 \\
   C_3 \\
  C_4 \\
   C_5 \\
   C_6 \\
 C_7
\end{array}
\right)  \rightarrow
\left( \begin{array}{l}
 1 \\
  0 \\
   0 \\
   0 \\
  0 \\
   0 \\
   0 \\
0
\end{array}\right)
   \label{examp1} 
\ee
Now have a Hadamard act on qubit 1 by use of the command 
$Op1[{\bf \cal{ H} },1,   \Psi_3].$ The result is displayed 
in vector form and then in Dirac form by use of the DForm 
command in Figure~\ref{OneHad}.
   
\begin{figure}
\fbox{\parbox{1.0\textwidth}{\includegraphics[width=14cm]{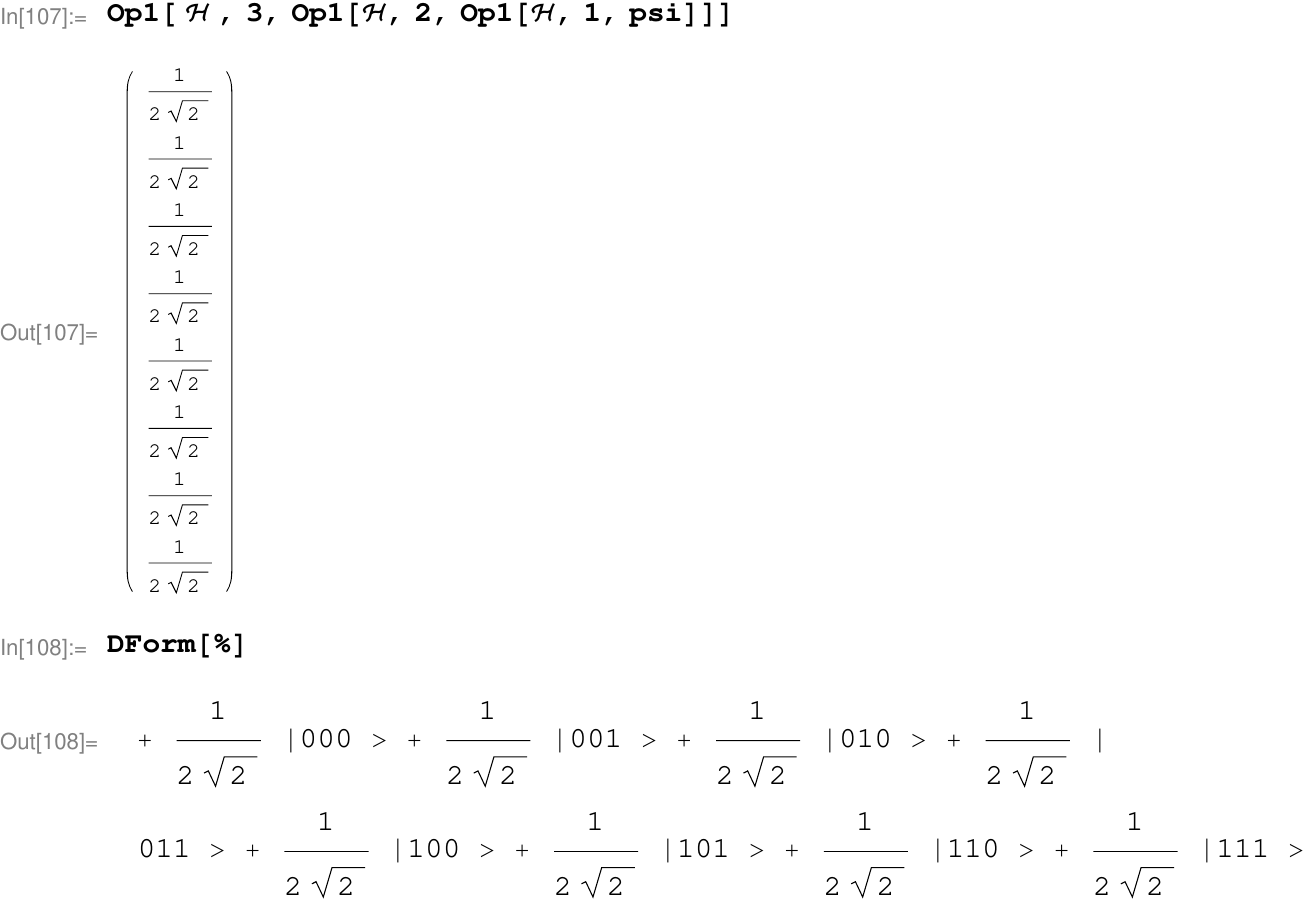}}}
\caption{Hadamards on all three qubits example. Here psi$= \mid 0 0 0 \rangle.$}
\protect\label{AllHad}
\end{figure}

\begin{figure}
\fbox{\parbox{1.0\textwidth}{\includegraphics[width=14cm]{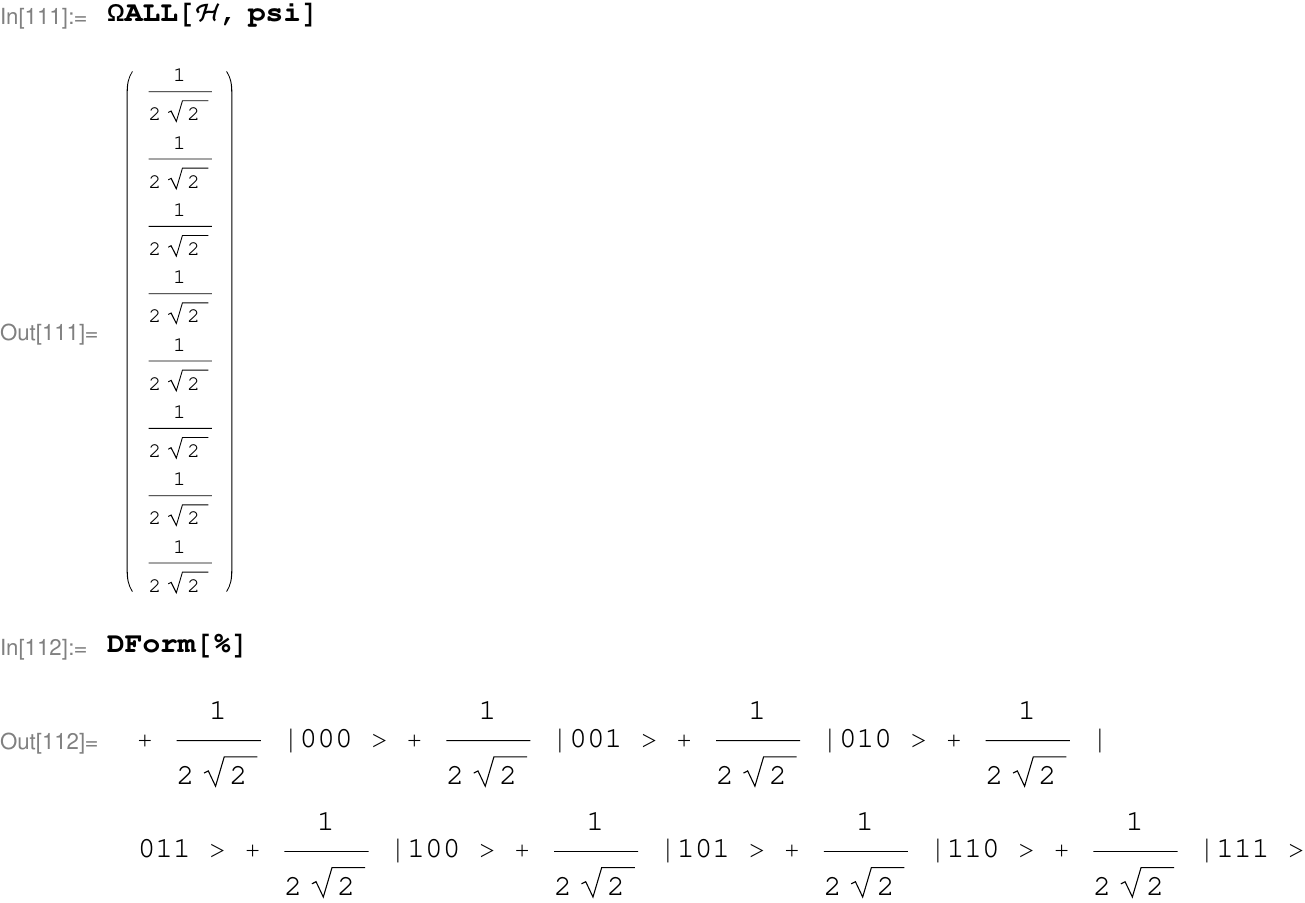}}}
\caption{Hadamards on all three qubits  using the $\Omega$ALL command. 
Here psi$= \mid 0 0 0 \rangle.$ }
\protect\label{AllHad2}
\end{figure}

\begin{figure}
\fbox{\parbox{1.0\textwidth}{\includegraphics[width=14cm]{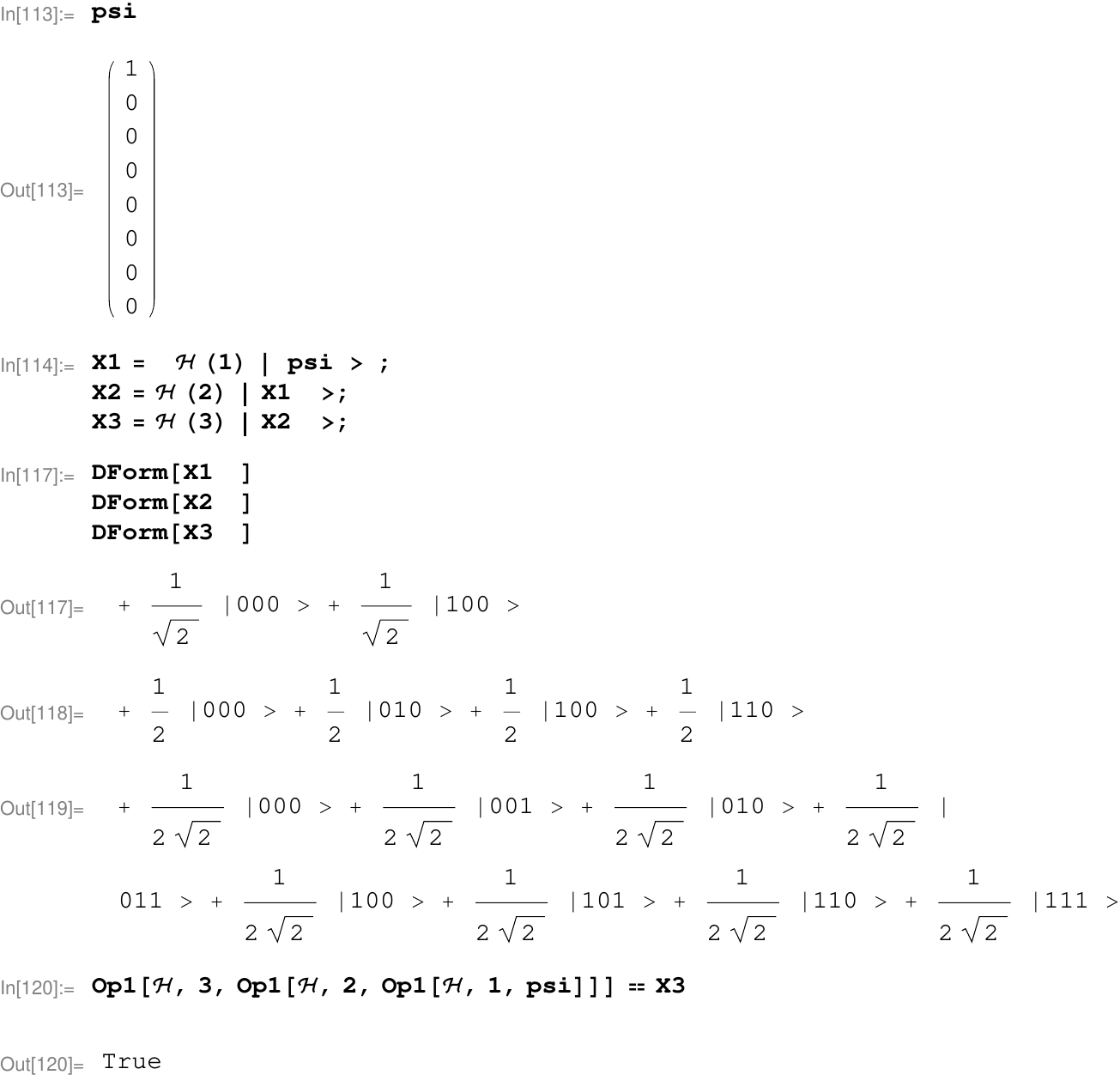}}}
\caption{Application of Hadamards in Dirac form.  Here psi$= \mid 0 0 0 \rangle.$}
\protect\label{Diracform1}
\end{figure}

One can act with Hadamards on every qubit, by either repeated 
use of Op1 Figure~\ref{AllHad}, or by the command 
$\Omega$ALL[${\bf \cal{ H}}$,psi],  which is illustrated in 
Figure~\ref{AllHad2}. A Dirac type notation is also available as 
illustrated in Figure~\ref{Diracform1}
  
\begin{figure}
\begin{center}
\fbox{\parbox{.7\textwidth}{\includegraphics[width=1.\textwidth]{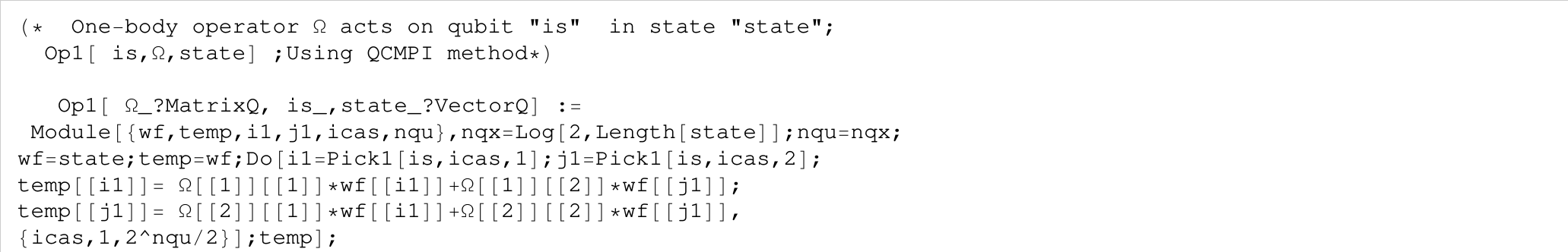}}}
\caption{The Op1 command as stipulated in {\bf QCWave.m}}
\protect\label{Op1command}
\end{center}
\end{figure}

In {\bf QCWave.m} the command Op1 is given as a Module see Figure~\ref{Op1command}, 
which makes use of the command Pick1, Pick1 selects the pairs of decimal 
labels which differ only in the ``is'' qubits value of 1 and 0. Then all such 
pairs are swept through.  Examples of Pick1 are presented in the Tutorial.
 
\subsection{Two-qubit operators}
\label{sec4b}

\begin{figure}[t]
\begin{center}
\fbox{\parbox{.7\textwidth}{\includegraphics[width=.7\textwidth]{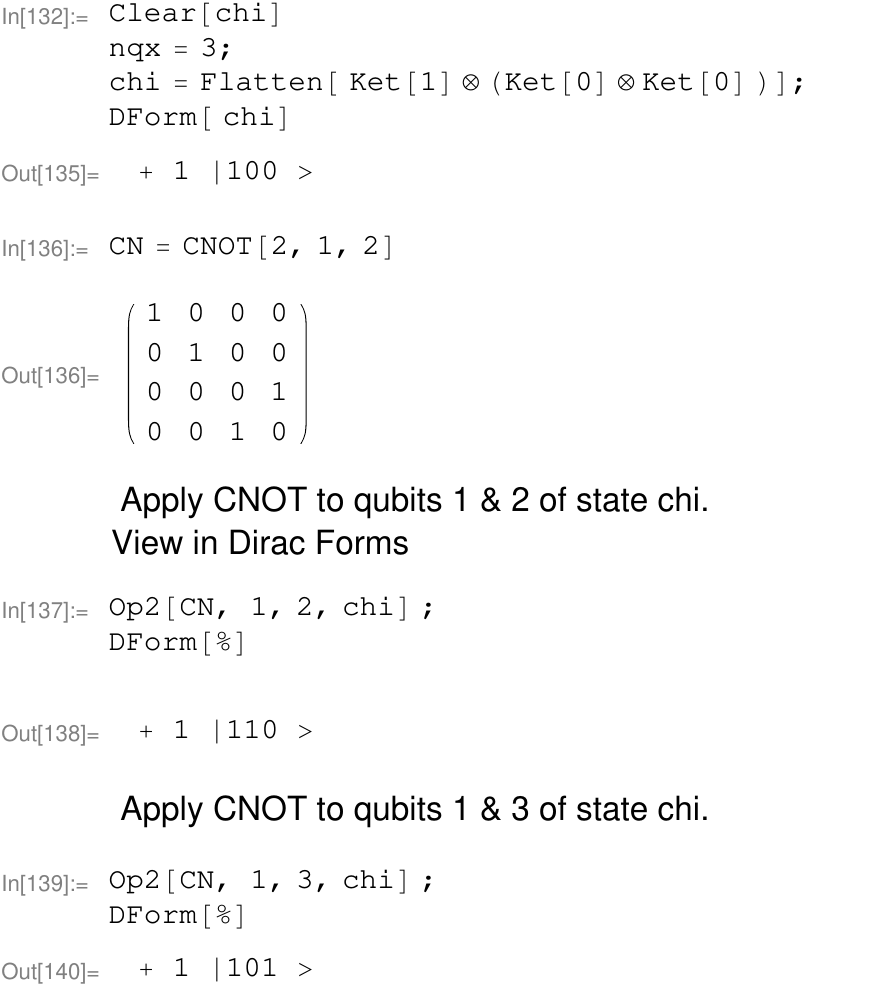}}}
\caption{Use of Op2 to apply CNOT gates.}
\protect\label{CNOTcommand}
\end{center}
\end{figure}

The typical two-qubit operators are the CNOT, and controlled phase 
operators. General two-qubit operators can be constructed from 
tensor products of two Pauli operators, as discussed earlier. An example 
from QCWave of application of a CNOT gate is presented in 
Figure~\ref{CNOTcommand}.

\begin{figure}
\begin{center}
\fbox{\parbox{.7\textwidth}{\includegraphics[width=.7\textwidth]{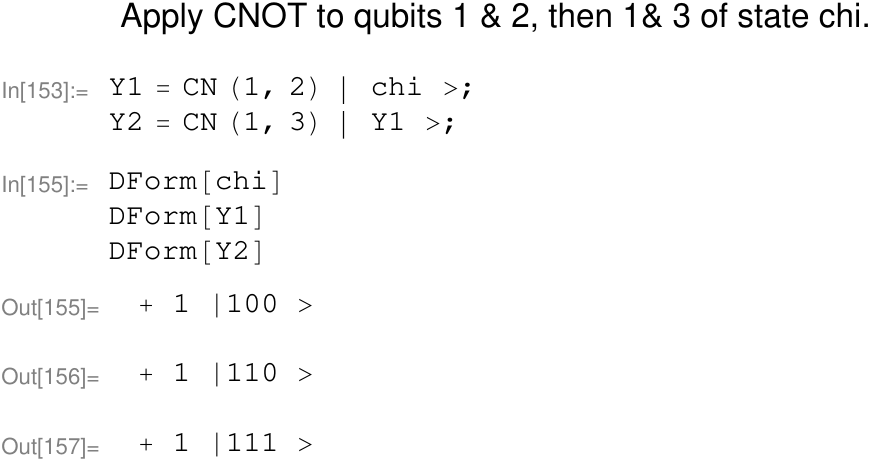}}}
 \caption{ Application of CNOT gates in Dirac form. }
\protect\label{Diracform2}
\end{center}
\end{figure}

A Dirac type notation is also available as illustrated in Figure~\ref{Diracform2}

\begin{figure}
\begin{center}
\fbox{\parbox{.9\textwidth}{\includegraphics[width=0.9\textwidth]{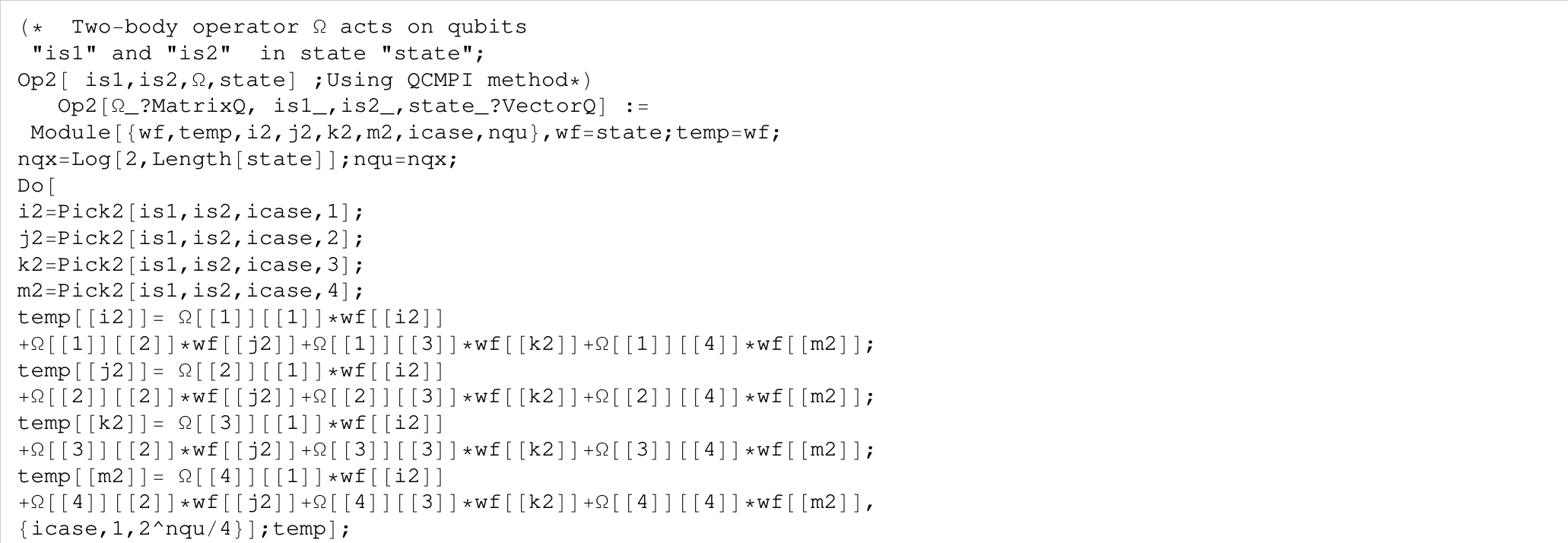}}}
 \caption{ The Op2 command as stipulated in {\bf QCWave.m}.}
\protect\label{Op2command}
\end{center}
\end{figure}

In {\bf QCWave.m} the command Op2 is given as a Module see Figure~\ref{Op2command}, 
which makes use of the command Pick2, Pick2 selects the quartet of decimal 
labels which differ only in the "is1" and "is2" qubit's values
of 1 and 0.  Then all such quartets are swept through. Examples of 
Pick2 are presented in the Tutorial.

\subsection{Three-qubit operators}
\label{sec4c}
The Op3 command is also provided in {\bf QCWave.m}  as a Module, which 
makes use of the command Pick3, Pick3 selects the octet of decimal 
labels which differ only in the "is1," "is2," and "is3" qubit's 
values of 1 and 0. Then all such octets are swept through. Application 
to the Tofolli gate is provided in the Tutorial.

\clearpage

\section{THE MULTIVERSE APPROACH}
\label{sec5}
\subsection{General remarks}
Mathematica 7.0 \& 8.0 provide a master-slave parallel processing 
facility. This is not a full implementation of a parallel processing 
setup that allows communication between the ``slave" processors'',  such 
as used by the MPI~\cite{MPI} protocol.  If MPI were readily available 
in Mathematica then one could invoke the full capabilities discussed in 
QCMPI. That capability allows for the state vector to be distributed 
over several processors which increases the number of qubits that could 
be simulated.  It is indeed possible to have Mathematica upgraded to 
include MPI slave to slave communication; as is available in the 
``POOCH''~\cite{pooch} package.  However, since that is an expensive 
route and most Mathematica users do not have access to MPI, although 
one can hope for such a capability in the future, we have not invoked 
the full state distribution aspect.

Nevertheless, the master-slave Mathematica 7.0 capability does provide 
for concurrent versions of a QCWave based algorithm to be run with 
different random error scenarios. Then an ensemble averaged density 
matrix can be formed which describes a real error prone QC setup. That 
opens the possibility of examining the role of errors and the efficacy 
of error correction methods using Mathematica.

Therefore, we provide a sample of a parallel setup using some simple basic
algorithms, where the parallel setup is described and explained in detail. 
The following steps are needed: (1) set up your Mathematica code to access 
several processors, see Appendix 1 for some help; (2) identify the processor 
number; (3)  introduce random errors depending on the processor number; 
(4) assign a probability distribution for the various processors; (5) form 
an ensemble average over the processors and store that information as a 
density matrix on the master processor;  (6) repeat these steps including, 
the algorithm, noise and finally error correction (EC) steps on all 
processors; (7) examine the resultant density matrix and its evolution to 
test the EC efficacy. This is an important program that we start by 
providing simple examples.

A quantum system can evolve in many ways.  Different dynamical evolutions 
are called paths~\cite{Feyn} or histories~\cite{griffiths}.  We refer to 
these alternate evolutions as separate ``universes'' and a collection 
of such possibilities as a multiverse or ensemble of paths. Parallel 
processing provides a convenient method for describing such alternate 
paths.

\subsection{The ideal and the noisy channels}

In our application,  we assume that the main path follows an ideal 
algorithm exactly and the alternate paths incorporate the algorithm 
with possible noise.  That noise is described by random one-qubit 
operators acting once, or with less likelihood twice.  To describe 
this idea, which is realized in the notebooks {\bf MV1-Noise.nb}, {\bf
  MV2-Noise.nb} and {\bf MVn-Noise.nb}, consider an initial density 
matrix $\rho_0.$  For a pure state, 
$\rho_0=\mid \psi_0\rangle  \langle \psi_0\mid ,$ but it can be a 
general initial density matrix subject only to the conditions 
$\rho^\dagger=\rho$ and $Tr[ \rho ]=1.$  The density matrix has 
$2^{2 n_q} -1$ parameters and $2^{n_q}$ real eigenvalues $\lambda_n \leq 1$ 
with $ \sum_n \lambda_n =1.$    The simplest one-qubit case has the form 
$ \rho = \frac{1}{2} ( 1 + \vec{P} \cdot \vec{\sigma} ),$  where the 
real polarization vector $ \vec{P} = Tr [ \vec{\sigma} \rho],$ is 
within the ``Bloch sphere'', $ (\vec{P} \cdot  \vec{P}) \leq 1.$~\footnote{The 
two qubit case is of the form 
$ \rho = \frac{1}{4} ( 1 + \vec{P_1} \cdot ( \vec{\sigma} \otimes  \mathbf{1})   +  
\vec{P_2} \cdot ( \mathbf{1}  \otimes  \vec{\sigma})  +         
\overleftrightarrow{\cal C} \cdot (   \overleftarrow{\sigma} \otimes   \overrightarrow{\sigma} )
), $  where  $ \vec{P_1}= Tr[  \vec{\sigma} \otimes  \mathbf{1})]\  \&\ 
 \vec{P_2} = Tr[   \mathbf{1}  \otimes  \vec{\sigma}   ]$ are the 
polarization vectors for qubits 1 and 2 and 
$ \overleftrightarrow{\cal C}=Tr[  \overleftarrow{\sigma} 
\otimes   \overrightarrow{\sigma}]$ is the $3 \times 3$ spin correlation 
tensor. Note the number of polarization plus correlations are 
$2^{2 n_q}-1$ = 3 (for one-qubit) and 15 (for two-qubits).  }

\subsubsection{Storage case}

Consider a simple case where the ideal algorithm is simply leaving 
the state, as described by $\rho_0,$ alone.  This is a memory storage 
case.  Ideally, $\rho$ remains fixed in time. Assume however that 
this ideal case occurs with a probability $\texttt{p}   \lesssim 1$  
and that alternate evolutions occur with a probability 
$\epsilon = 1- \texttt{p} .$ For example, we take $\texttt{p}=.8$ 
and $\epsilon =.2,$ corresponding to a 80\%  perfect storage and 
20\% possibility of noise. We also assume for more than 1 qubit 
cases that 95\% of the 20\% noise ($.2 \times .95\rightarrow 19\% $) 
involves a single one-qubit hit, while 5\% of the 20\% noise 
($.2 \times .05\rightarrow 1\% $) involves two one-qubit hits.

The ensemble average over all paths then yields a density matrix
\be
\rho_f =\texttt{p}\   \rho_0 
+\frac{ \epsilon}{n_p} ( \Omega^{(1)}\  \rho_o\  \Omega^{(1)\dagger }
+\Omega^{(2)} \ \rho_o \ \Omega^{(2)\dagger } \cdots  \Omega^{(n_p)}\  \rho_o
\  \Omega^{(n_p)\dagger }),
\label{rhoE1} 
\ee   
where the operators  $ \Omega^{(1)},  \Omega^{(2)} \cdots  \Omega^{(n_p)}$ 
act in each of the $n_p$ paths with a probability $ \frac{ \epsilon}{n_p}. $  
Each of these $n_p$ terms is evaluated on a separate processor, so that 
$n_p$ equals the total number of processors invoked. The above ensemble 
average preserves the trace:
\be
Tr[ \rho_f ]= \texttt{p}\   Tr[ \rho_0 ] +\frac{ \epsilon}{n_p}  
\sum_{n=1}^{n=n_p}{Tr[\Omega^{(n)}\  \rho_o\  \Omega^{(n)\dagger ] }}
=  \texttt{p} + \frac{ \epsilon}{n_p} n_p =1.
\ee 
Here we assume that each $ \Omega^{(n)\dagger} \Omega^{(n)}  =1,$ and hence 
that $ Tr[\Omega^{(n)}\  \rho_o\  \Omega^{(n)\dagger} ] =Tr[  \rho_o\ ] =1.$ 
In addition, $\rho_f^\dagger = \rho_f.$
  
\begin{figure}
\begin{center}
    \setlength{\unitlength}{1in}
      \begin{picture}(1,2)
      \thicklines
     \bezier{25}(0,0)(0.5,.9)(1.8,0.7)
      \bezier{25}(0,.4)(0.5,1.3)(1.8,1.1)
      \bezier{50}(0,.8)(0.5,1.7)(1.8,1.5)
      \bezier{300}(0,1)(0.5,1.9)(1.8,1.7)
      \put(-0.18,+.9){\makebox(.2,.2)[tl]{$t_0$} }
       \put(1.55,1.7){\makebox(.4,.4)[br]{t} }
       \put(-0.3,+.65){\makebox(.2,.2)[tl]{n=1} }
        \put(-0.3,+.3){\makebox(.2,.2)[tl]{n=2} }
         \put(-0.3,-.15){\makebox(.2,.2)[tl]{n=$n_p$} }
       \put(+0.5,+1.25){\makebox(.2,.2)[tl]{$\Omega^{(1)}$} }
        \put(+0.5,+.85){\makebox(.2,.2)[tl]{$\Omega^{(2)}$} }
         \put(+0.5,.45){\makebox(.2,.2)[tl]{$\Omega^{(n_p)}$} }
        \put(+0.95,+1.4){\makebox(.2,.2)[tl]{$\tilde{\Omega}^{(1)}$} }
        \put(+0.95,+1.0){\makebox(.2,.2)[tl]{$\tilde{\Omega}^{(2)}$} }
         \put(+0.95,.6){\makebox(.2,.2)[tl]{$\tilde{\Omega}^{(n_p)}$} } 
       \put(-0.6,.45){\makebox(.2,.2)[tl]{$\rho_0$} }
       \put(2.0,.95){\makebox(.2,.2)[tl]{$\rho_F$} }
         \put(+0.5,+1.7){\makebox(.2,.2)[tl]{$\Omega_A$} }
          \put(+0.9,+1.75){\makebox(.2,.2)[tl]{$\tilde{\Omega}_A$} }
   \end{picture}
\end{center}
\caption{The multiverse approach. Each curve represents a possible 
evolution of a quantum system of $n_q$ qubits. The solid curve represents 
the main evolution path, with probability $\texttt{p}.$  The dotted 
curves represent the $n^{th}$ evolution path, with probability 
$\frac{\epsilon}{n_p}.$ The initial density matrix is $\rho_0$ and 
the final $\rho_F.$ The unitary noise operators $\Omega^{(i)}$ act on 
each path $i.$ Here $\tilde{\Omega}^{(i)}$ denote subsequent noise 
operators. The operators $\Omega_A$ and $\tilde{\Omega}_A$ denote the 
$n_q$ algorithm operators which act on all paths.}
\protect\label{scheme}
\end{figure}
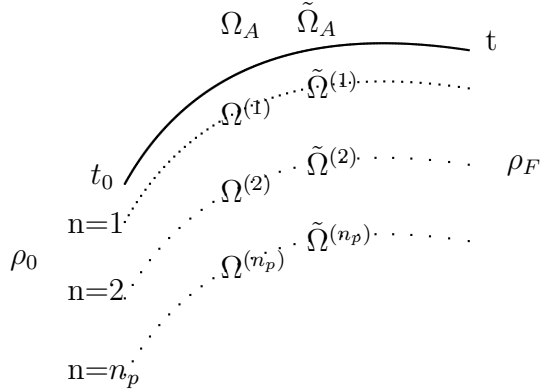

This multiuniverse approach is illustrated in Figure~\ref{scheme}. For 
the pure storage case the algorithm operators $\Omega_A, \tilde{\Omega}_A
\cdots$ are all set equal to unit operators. See later for a simple 
non-trivial case.

\subsubsection{Multiverse and POVM}

The above representation can also be cast in the POVM (Positive Operator 
Valued Measure) and Kraus operator form. The evolution can be expressed 
as
\be
\rho_f = \sum_{n=0}^{n=n_p} \   
\tilde{\Omega^{n}  } \  \rho_0 \   \tilde{\Omega}^{n \dagger},  
\label{rhoE2} 
\ee 
where we define $\tilde{\Omega}^{0} = \sqrt{ \texttt{p}} \  \mathbf{1} ,$ 
and $\tilde{\Omega}^{n\rangle0} = \sqrt{ \frac{\epsilon}{n_p}}  \Omega^{n}.$ 
This is the form known as POVM, which can be deduced~\cite{Preskill} 
from embedding a quantum system in an environment, which is then projected 
out.  This evolution form can also be used to deduce the
Lindblad~\cite{Lindblad} equation for the evolution of a density matrix 
subject to environmental interactions. Here we arrive at these forms 
from a simple multiuniverse approach.
    
\subsubsection{Multiverse and classical limit}
 
Our task is to set up this multiuniverse approach using the 
parallel, multi-processor features of Mathematica. Equation~\ref{rhoE1} 
describes the evolution of a density matrix after one set of 
operators act in the various possible paths. A subsequent set 
of operators is described by
\be
\rho_F =\texttt{p}\   \rho_f +\frac{ \epsilon}{n_p} \sum_{n=1}^{n=n_p}  ( \caret{\Omega}^{(n)}\  \rho_f\  \Omega^{(n)\dagger }).
\label{rhoE3} 
\ee   
As this evolution process continues to be subject to additional noise 
operators, the density matrix evolves into a diagonal or classical form.  
In this way the noise yields a final classical density matrix with 
zero off-diagonal terms;  this is the decoherence caused by a quantum 
system interacting with an environment. If the qubit states are not 
degenerate and the noise is of thermal distribution, the density matrix 
in the classical limit will evolve towards the thermodynamic form 
$\exp(-\frac{H}{k T}).$ At every stage, one can track the von Neumann 
entropy( $S(\rho) = - Tr[\rho \ln[\rho]] ),$ the Purity 
( $Tr[\rho \cdot \rho]),$ and the Fidelity ( $F[\rho,\rho_0]= Tr[\sqrt{ \sqrt{\rho_0} \  
  \cdot  \rho \cdot\  \sqrt{\rho_0}  }  ]  ),$~\footnote{To evaluate this 
complicated expression, we find the eigenvalues of $\rho. \rho_0 $ and 
then form the sum $\sum_{n}  \sqrt{ \mid \tilde{\lambda}_i\mid},$ to obtain 
a good approximate value.} of the system. In addition, the eigenvalues 
of the system can be monitored where in the classical limit the eigenvalues 
all approach  $ \frac{1}{2^{n_q}},$ and the entropy goes to 
$S[\rho]\rightarrow n_q.$  Subsystem entropy and eigenvalues can also 
be examined.
   
\subsection{Multiverse algorithms and errors}
   
The evolution of the density matrix, with noise included via the multiverse 
approach on the available processors, can also be implemented when an 
algorithm is included. The procedure is to act with the algorithm gate 
operators after each ensemble averaged density matrix is formed. The explicit 
expression is given in Equation~\ref{rhoE1} which for the $nth$ step is
\be
\rho_{n+1} = \Omega_A \   \left[   \rho_n \    + \frac{\epsilon}{n_p}\ 
\sum_{k=1}^{k= n_p}\ \   \sum_{s=1,2} \  p_s \Omega_{k s}  \rho_n  \Omega_{k s}^\dagger\  \right]  \ \Omega_A^\dagger
\label{rhoE4} 
\ee  
where $\Omega_A$ are the gates for the specific algorithm and 
$\Omega_{k s}$ are the noise operators on the $kth$ processor for 
two cases, $s=1$ denotes a one qubit noise operator hitting one 
qubit and $s=2$ denotes one qubit operators hitting two separate 
qubits. The factors $p_k$ are assumed to be $p_1=.95$ and $p_2=.05,$ 
so that the one-qubit hits have a net probability ($\epsilon*p_k$) 
of 19\% and the double hit case a net probability of 1\%. The 
algorithm operators $\Omega_A$ are applied to all all processors, 
but implemented by evaluation on the master processor.
   
\begin{figure}[h]
\fbox{\parbox{1.0\textwidth}{\includegraphics[width=14cm]{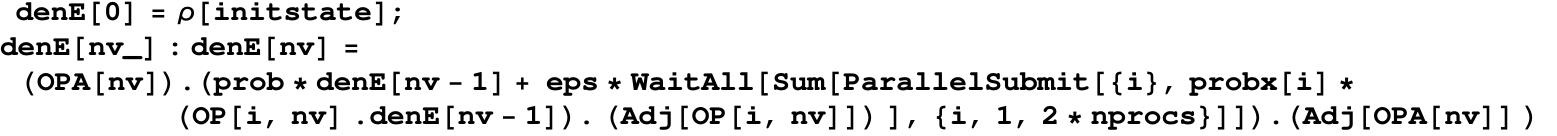}}}
\caption{Here denE[0[] is the initial density matrix $\rho_0$ and 
denE[nv] denotes the ensemble averaged density matrix after nv steps. 
The algorithm operators  ``OPA[nv]'' are setup within the code, 
and the noise operators OP[ i , nv ] act on the $ i^{th} $ processor 
at the $ nv^{th} $ step. These noise operators are held fixed after 
they are randomly generated. }
\protect\label{denE}
\end{figure}

In QCWAVE, the above steps are implemented in the notebooks {\bf MV1-Noise}, 
{\bf MV2-Noise} and {\bf MVn-Noise}, for systems consisting of 1,2, or 
$n$ qubits. The key step is shown in Figure~\ref{denE}, where denE[n] 
denotes the ensemble averaged density matrix at stage $n$, and the parallel 
part of the command distributes the evaluation of the noise over the 
``nprocs'' processors,  which is doubled to account for the ``s'' label 
in Equation~\ref{rhoE4}.   

\begin{figure}[th]
\begin{center}
{\includegraphics[{width=1.2\textwidth}]{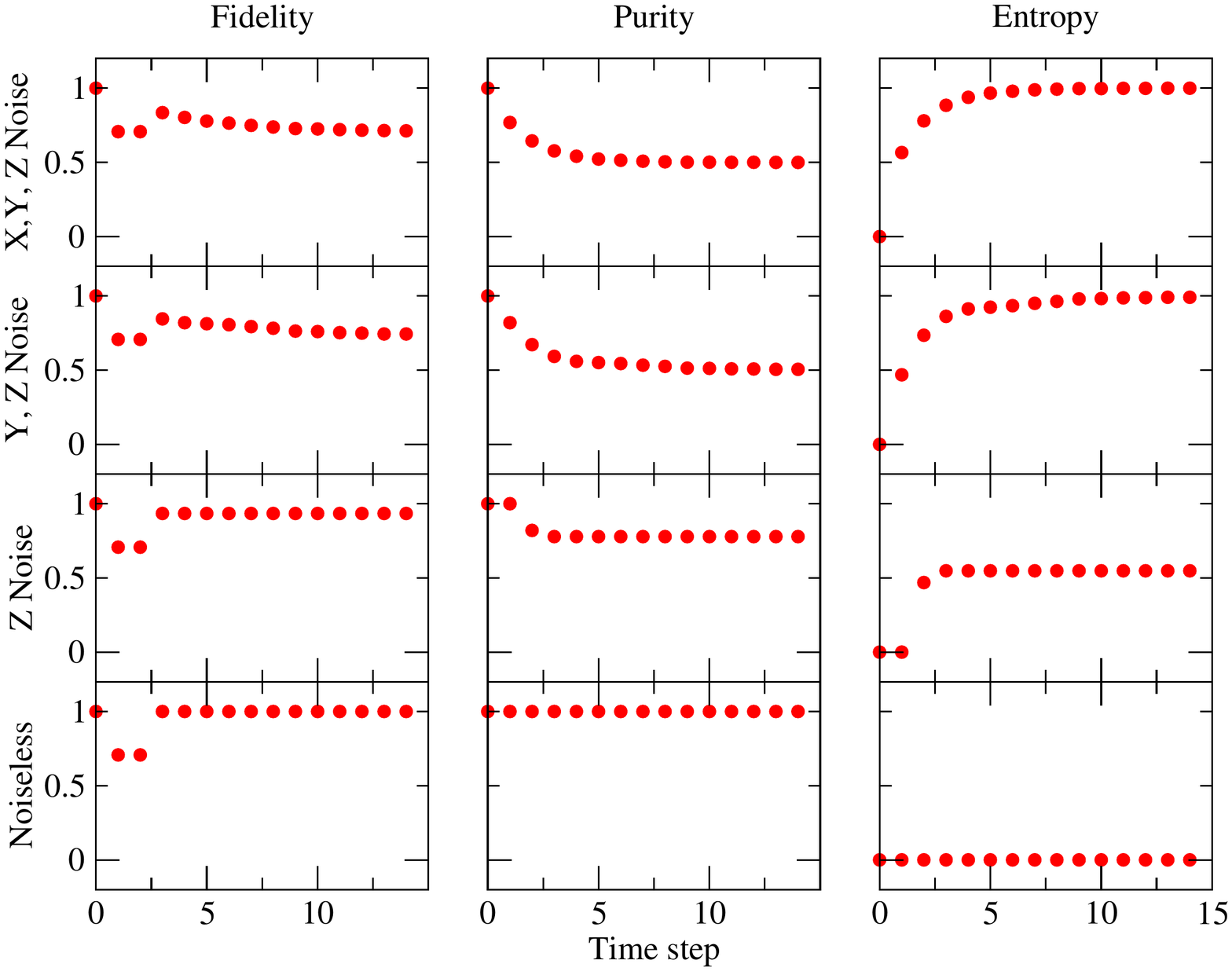}}
\end{center}
\caption{ The fidelity (left column), purity (central column) and 
entropy (right column)  evolution is displayed for one qubit initially 
in  a state $\mid \psi_0\rangle=\mid 0 \rangle,$ that is subject to 
a Hadamard at step 1 and another Hadamard at step 3, for various 
noise scenarios. The top plot is for X,Y and Z noise, the next plot 
down has Y and Z noise, and the next Z-noise only. The bottom plot 
is the noiseless channel, for which the fidelity returns to one, 
after dropping to $1/\sqrt{2},$  which clearly reflects the action 
of 2 sequential Hadamards. }
\protect\label{MV1FID}
\end{figure}

A simple algorithm is illustrated in {\bf MV1-Noise}; namely, one 
starts with the state $\mid 0\rangle$ which is then hit by a 
Hadamard ${\bf \cal{ H} },$ and after an interlude of noise, 
another Hadamard hits, followed by a long sequence of noise.
Without noise this process correspond to a rotation to the 
x-axis and then a rotation back to the z-axis. One also sees 
the polarization vector rotated, rotated back and then, after 
a sequence of noise hits, decay to zero and density matrix 
then evolves into a diagonal form, with both eigenvalues equal 
to 1/2. How is this simple process affected by noise during 
these steps?  To answer that question the entropy, purity and 
fidelity evolution are tracked. The results from {\bf MV1-Noise} 
are illustrated in Figure~\ref{MV1FID}.

\begin{figure}[t]
\begin{center}
{\includegraphics[{width=1.2\textwidth}]{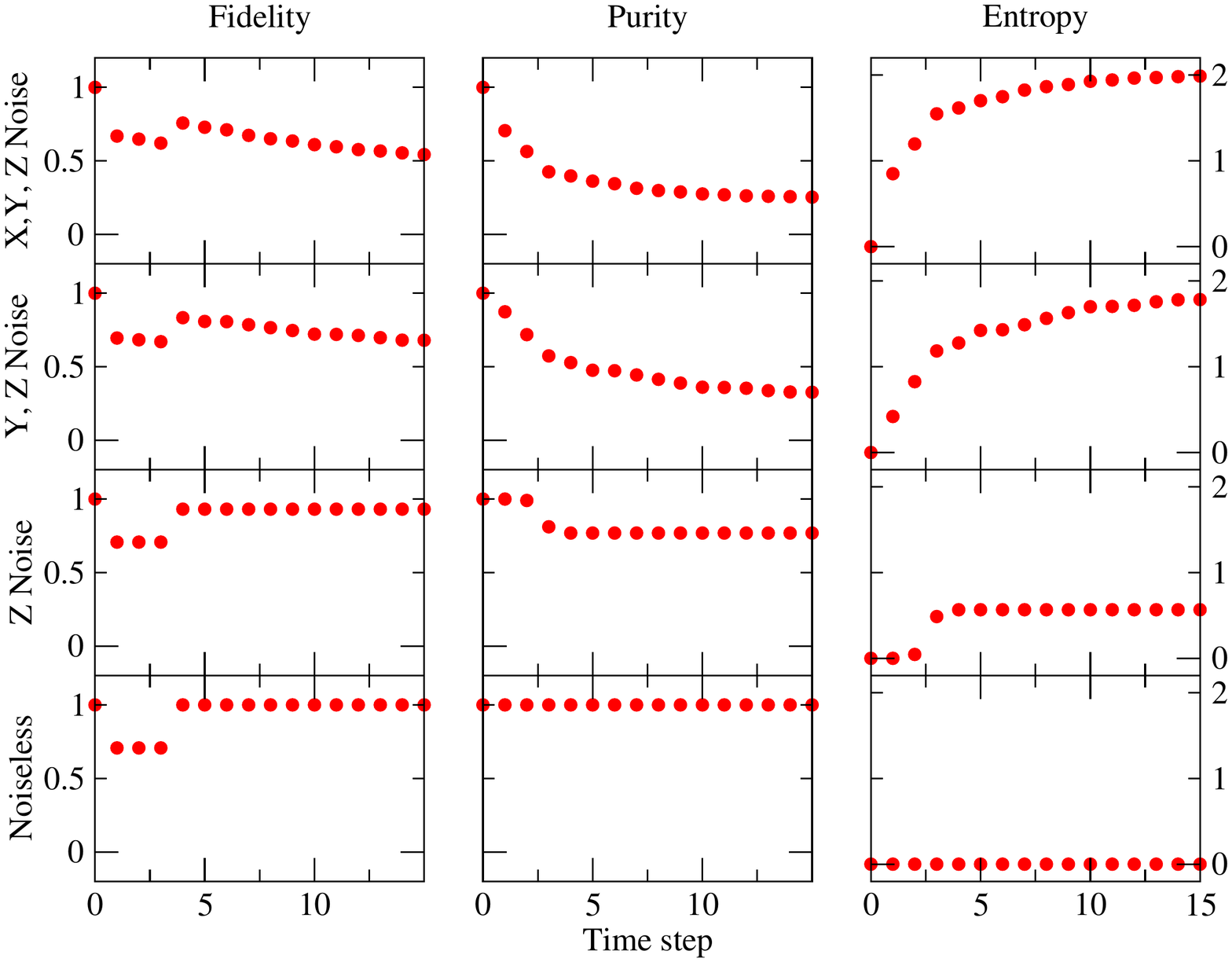}}
\end{center}
\caption{The fidelity (left column), purity (central column) and 
entropy (right column) evolution is displayed for two qubits initially 
in a state $\mid \psi_0\rangle=\mid 0 0 \rangle,$ that is subject to 
a CNOT$_{1 2}{\cal H}_1$  and later a ${\cal H}_1 $CNOT$_{1 2}$, for 
various noise scenarios. The top plot is for X,Y and Z noise, the 
next plot down has Y and Z noise, and the next Z-noise only. 
The bottom plot is the noiseless channel, for which the fidelity 
returns to one, after dropping to $1/\sqrt{2},$  which clearly reflects 
the action of the 2 sequential operators.}
\protect\label{MV2FID}
\end{figure}

Another simple algorithm is illustrated {\bf in MV2-Noise}; namely, 
one starts with the state $\mid 0 0\rangle ,$ qubit one is then 
hit by a Hadamard  ${\bf \cal{ H} },$ and after an interlude of 
noise, a CNOT gate acts on both qubits.  This is the algorithm for 
producing a Bell state CNOT$_{1,2} { \cal H}_1 \mid 00\rangle 
= \frac{1}{2} ( \mid 0 0 \rangle + \mid 1 1 \rangle  ).$ This is 
followed by an inverse Bell operator ${ \cal H}_1 $CNOT$_{1,2}, $ and 
then a long sequence of noise. The  density matrix then evolves 
into a diagonal form, with all 4 eigenvalues equal to 1/4. The two 
polarizations, and the spin correlations are displayed along with 
the evolution of the eigenvalues, the entropy, purity and fidelity. 
The results from {\bf MV2-Noise} are illustrated in 
Figures~\ref{MV2FID}.--~\ref{MV2ENT}.
   
Clearly, more sophisticated algorithms can be invoked. We next 
consider how to monitor and correct for the noise.
   
\subsection{Multiverse and error correction}
   
\subsubsection{Simulation of error correction}
   
Error correction (EC) typically involves encoding the qubits using 
extra qubits, then entangling those encoded qubits with auxiliary 
qubits. Measurements are made on the auxiliary qubits, so as not 
to disturb the original encoded qubits. Those measurements provide 
information as to whether an error has occurred, its nature and 
where it acted. Hence a remedial gate can be applied to undo the 
error. If desired, the encoded qubits can then be decoded and the 
original error-free qubit restored.  That process is illustrated 
for simple X and Y errors on one qubit EC and for Shor's 9 qubit 
EC code in notebooks  {\bf EC3x, EC3z and Shor9Tutorial}.  More 
sophisticated EC codes are available in the literature, along 
with a general theoretical framework~\cite{Gottesman}. This kind 
of EC has to be constantly invoked as an algorithm evolves, which 
is a rather awkward and qubit-costly process.  Error in the gates 
themselves is an additional concern, usually one assumes perfect 
gates, with errors (noise) occurring only in-between application 
of the gates.
   
For our purpose, instead of applying the procedures outlined in 
the above EC notebooks, we simulate EC by a rather simple procedure. 
In the notebooks, the operators ss[i] are set equal to the Pauli 
operators s[i] to generate noise. By replacing ss[1] by s[0] 
(the $2\times2$ unit operator), all X-noise is turned off ``by hand.''  
Then a rerun is generated which has the same structure as with the 
noise, except the X-noise has been removed.  Similar steps can be 
used to remove the Y and the Z-noise operators. In that way a set 
of results can be generated ranging from a full noise, to partial 
noise to no noise cases. Examples from {\bf MV1-noise} and {\bf MV2-noise} 
are presented in Figures~\ref{MV1FID}
and \ref{MV2FID}.

\begin{figure}[t]
{\fbox{\includegraphics[width=\textwidth]{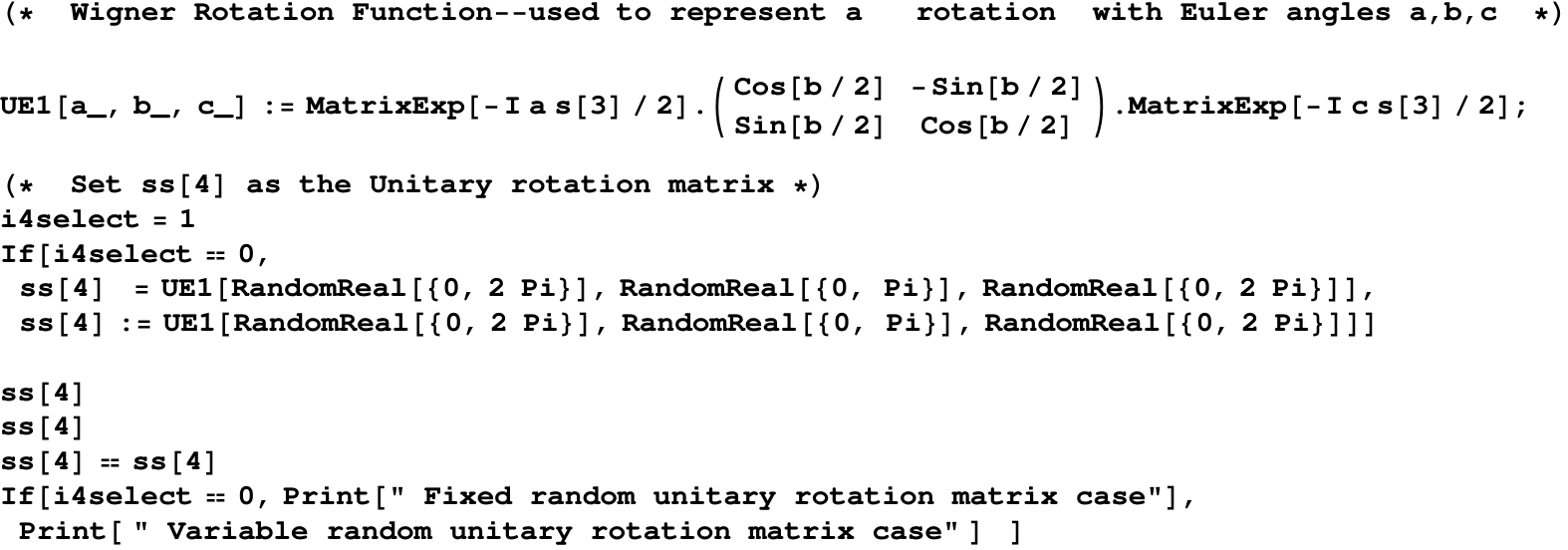}}}
\caption{ This is an option to use a general rotation as the 
noise operator as an alternative to the Pauli operators. The s[4] 
location can be used to include this option.}
\protect\label{Gnoise}
\end{figure}

If the user wishes to invoke other noise operators, that can be 
accommodated as well. For example, a general unitary random rotation 
can be used as a noise operator by invoking the form shown in 
Figure~\ref{Gnoise}. Thus "ss[4] =UE1"  is used to turn on such 
rotations. Replacing s[4] to s[0] again provides a way to turn this 
operator off to remove that noise element.
      
With this simple scheme, one can study many more noise and EC scenarios. 
For example, in the notebook {\bf MVn-Noise} the case of an algorithm 
for 5 qubits is presented.  The algorithm consists of a Hadamard followed 
by a CNOT chain, i.e. CNOT$_{1 5}$ CNOT$_{1 4}$ CNOT$_{1 3}$ CNOT$_{1 2} {\cal H}_1
\mid 00 \rangle.,$ including noise in-between and after the 9th step. 
Detailed examination of the entropy, fidelity, purity and eigenvalues 
and noise is presented within that notebook. Of particular interest is 
the EC simulation results when the noise operators are turned of sequentially. 

\section{ADDITIONAL FEATURES}
\label{sec6}

\subsection{Amplitude displays}
\label{sec6a}
The amplitude coefficients $C_n$ can be displayed in various ways 
using the commands {\bf Amplitudes} and {\bf MeterGraph} as illustrated 
in Figure~\ref{amps}

\begin{figure}[h]
\fbox{\parbox{0.8\textwidth}{\includegraphics[width=0.8\textwidth]{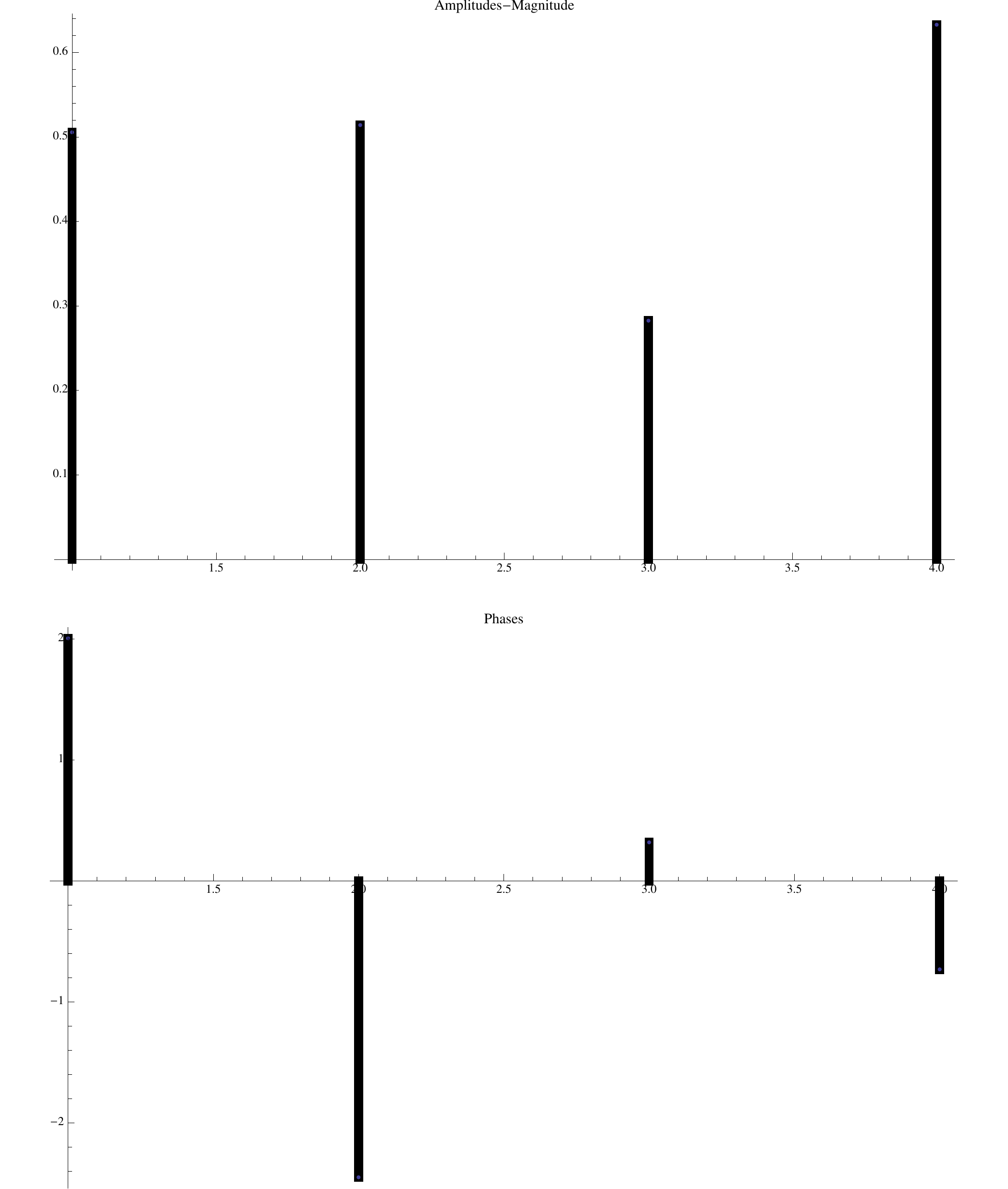}}}
\caption{ {\bf Amplitudes} 
  command displays amplitudes as magnitude and phase bar graphs.--see  {\bf QCWave.m} }
\protect\label{amps1}
\end{figure} 
\begin{figure}[h]
\fbox{\parbox{1.0\textwidth}{\includegraphics[width=14cm,height=4cm]{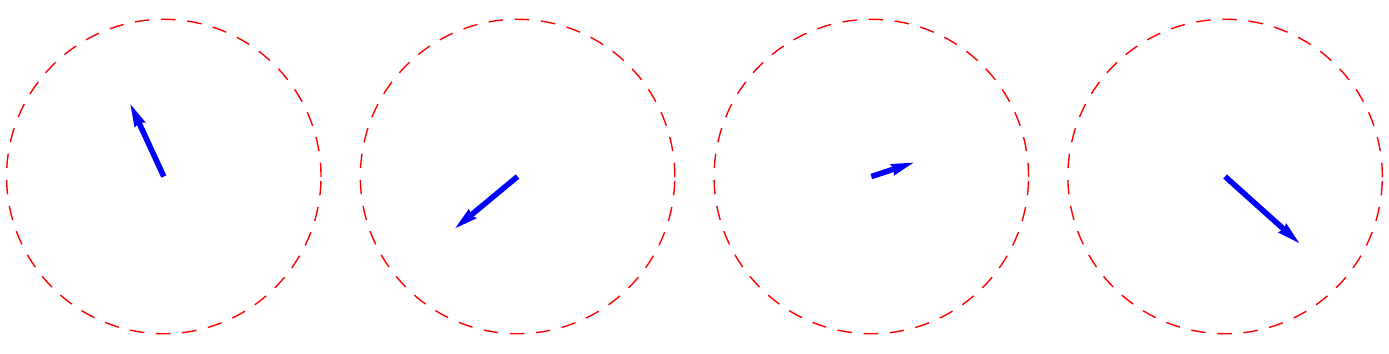}}}
\caption{  {\bf MeterGraph}
   displays amplitudes as argand plots--see  {\bf QCWave.m} }
\protect\label{amps2}
\end{figure}

\subsection{Dirac form}
\label{sec6b}

The command DForm has already been demonstrated in Figures 1--4,and 6. 
Another Dirac form has be invoked in QCWAVE as shown in Figure 4. A more 
extensive Dirac notation scheme has been provided by Jos\'e Luis
G\'omez-Mu\~noz  {\it et al.} in Ref.~\cite{mexico}.

\subsection{Circuit diagrams}
\label{sec6c}

\begin{figure}
{\includegraphics[{width=14cm,height=14cm}]{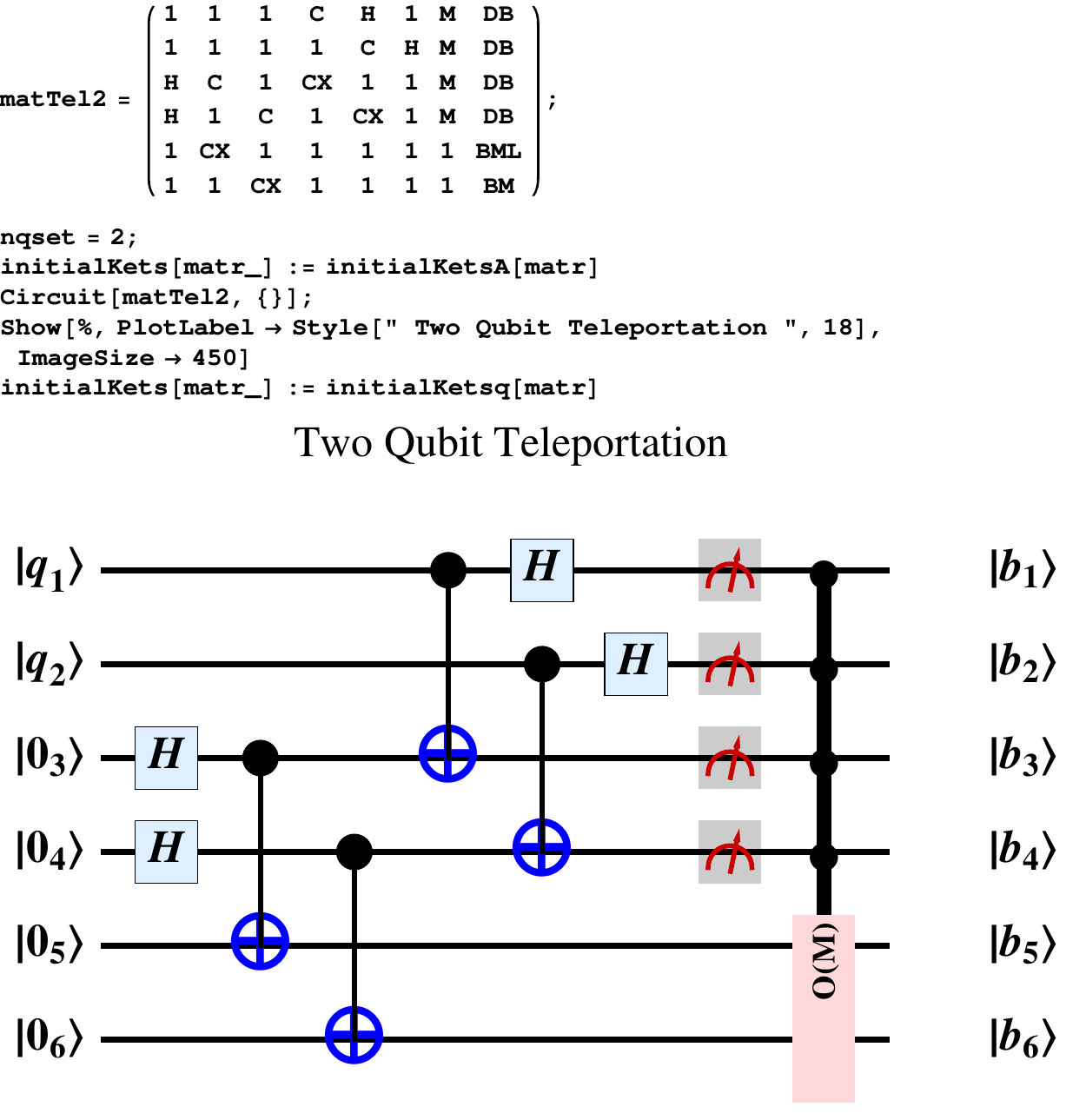}}
\caption{A sample circuit diagram as produced in notebook 
{\bf TeleportationW} is shown here. The initialKetsA sets 
the initial array sequence as: $\mid q_1\rangle, \mid q_2\rangle , \mid 0_3\rangle \cdots \mid 0_6\rangle.$}
\protect\label{circuitex}
\end{figure}

Illustrations of circuit drawing are included throughout the notebooks, 
with the {\bf CircuitTutorial} notebook providing an overview. 
The commands are all defined in  {\bf Circuits.m}.  One example is 
given in Figure~\ref{circuitex}.

\subsection{ Upgraded applications}
\label{sec6d}

Upgraded versions of Grover~\cite{Grover}, Teleportation~\cite{Teleportation} 
and Shor~\cite{Shor} algorithms are included in the present version.

\section{CONCLUSION \& FUTURE APPLICATIONS}

This package will hopefully be instructive and useful for applications to 
error correction studies. Hopefully users will contribute to improvements 
and extensions and for that purpose we are developing an interacting web
page. When MPI becomes available on Mathematica, there will be another 
opportunity to upgrade QCWAVE to a full research tool.

Application to explicit quantum computing problems, such as study of 
non-degenerate states and the associated phase factors, errors in gates 
themselves (where the gates are produced by explicit pulses), and the 
direct application of EC schemes are among the possible future applications. 
Novel EC schemes, such as stabilizing pulses or EC stable spaces 
could be additional fruitful applications.

\label{sec7}
\clearpage
\appendix
\section{Setting up Processors with Mathematica }

\begin{figure}[h]
\fbox{\parbox{1.0\textwidth}{\includegraphics[width=14cm]{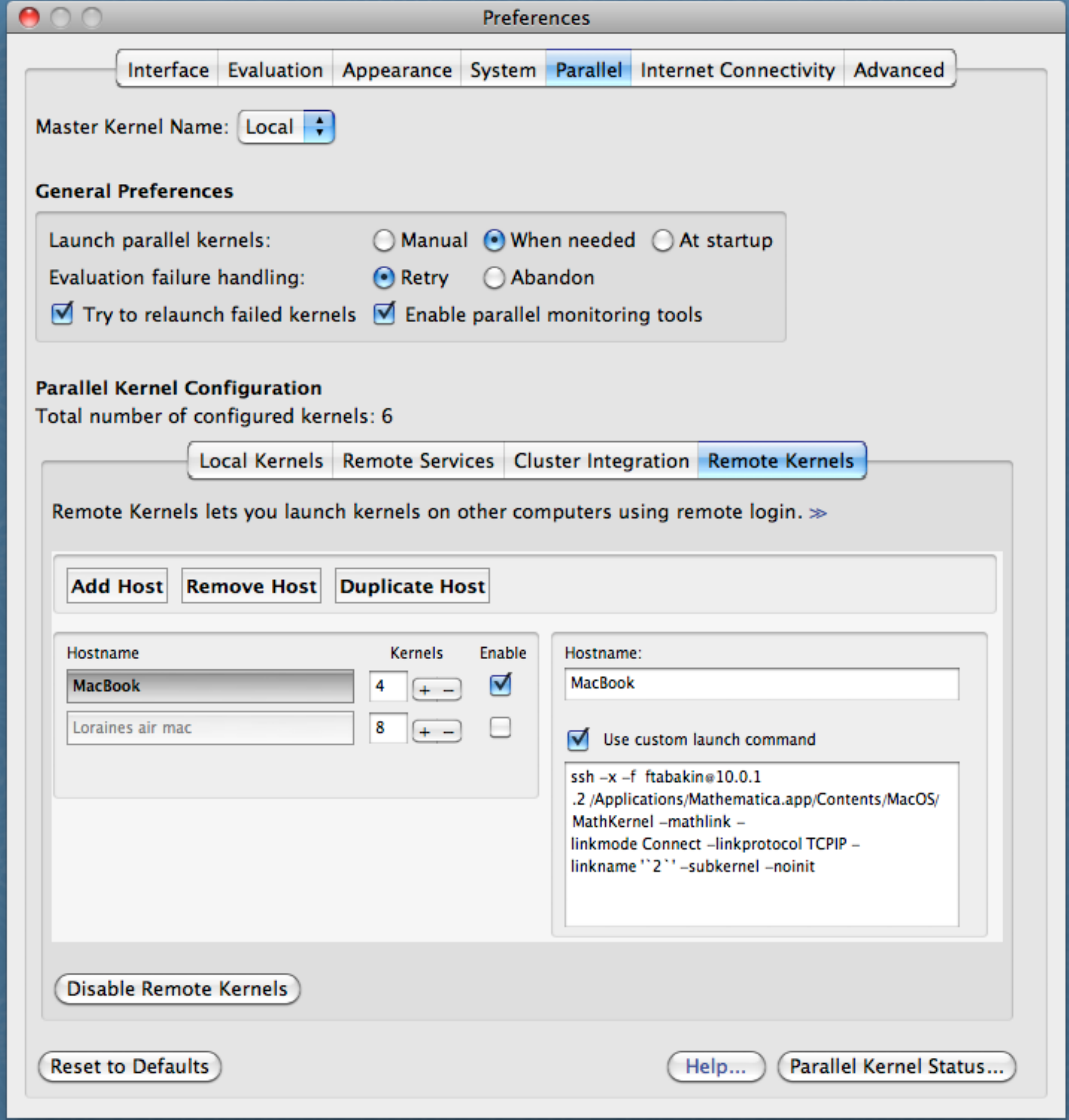}}}
\caption{The preferences setup to access other processors-- use 
``Evaluation/Parallel Kernel Configuration/Remote Kernels'' to get 
to this page. You also need to set the Local Kernels entry. The user 
also needs to establish an ssh link to the other machines and  be 
sure it properly accesses the Mathematica installed on all processors.}
\protect\label{setup}
\end{figure}

In order to use parallel processing with Mathematica, one needs to first 
gain access to several processors. There are other ways to do this, but, 
the commands we used are given in Figure~\ref{setup}. Note that the user
need to be sure that the ssh (secure shell) access is working and accesses 
the Mathematica command on the other machines (which could be Macs or 
PCs or a combination of them).

\begin{figure}[h]
\fbox{\parbox{.7\textwidth}{\includegraphics[width=.7\textwidth]{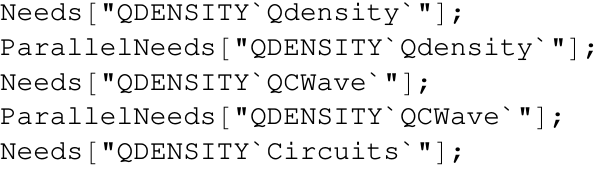}}}
 \caption{The commands needed to invoke the packages on parallel 
processors. {\bf QCWave.m and Circuits.m} are add-ons to the original 
{\bf QDensity.m} package. }
\protect\label{needs}
\end{figure}

In addition, one needs to setup the basic programs and requisite 
packages on all the processors used. For that the initializations 
shown in Figure~\ref{needs} are needed.

\section*{Acknowledgments}
This project was supported  earlier in part by the U.S. National 
Science Foundation and in part under  Grants  PHY070002P  \&  PHY070018N  
from the Pittsburgh Supercomputing Center, which is supported by several 
federal agencies, the Commonwealth of Pennsylvania and private industry. 
B.J-D. is supported by a CPAN CSD 2007-0042 contract. This work is also 
supported by Grants No. FIS2008-01661 (Spain), and No. 2009SGR1289 from 
Generalitat de Catalunya.

\end{document}